\documentclass{ar-1col}
\usepackage[numbers]{natbib}
\usepackage{amsmath}
\usepackage{bm,color,wrapfig,lipsum}
\usepackage{epstopdf}
\usepackage{url}
\usepackage{bbm}
\usepackage{dsfont}

\newcommand{\tBox}[1]{{Figure~\ref{box:#1}}}
\newcommand{\twoBoxes}[2]{{Figure~\ref{box:#1} and~\ref{box:#2}}}

\newcounter{jpsBoxcnt}
\renewcommand{\thejpsBoxcnt}{\Roman{jpsBoxcnt}}

\newenvironment{jpsBox}[3]
{
\begin{textbox}[#1]
\refstepcounter{jpsBoxcnt}

\section{Figure~\thejpsBoxcnt: #2}

\label{box:#3}

\addcontentsline{toc}{subsection}{\thejpsBoxcnt: #2}
}
{\end{textbox}}

\newcommand{\e}{{\mathrm{e}}}
\renewcommand{\d}[1]{{\mathrm{d}}#1}

\newcommand{\stress}{{\sigma}}
\newcommand{\critstress}{{\sigma_{\mathrm{c}}}}

\newcommand{\blue}[1]{{\color{blue}{#1}}}

\jname{Xxxx. Xxx. Xxx. Xxx.}
\jvol{AA}
\jyear{YYYY}
\doi{10.1146/((please add article doi))}

\begin{document}

\markboth{Sethna et al.}{Deformation of crystals}

\title{Deformation of crystals: Connections with statistical physics}

\author{%
\vskip -0.3cm
James P. Sethna,$^1$ 
Matthew K. Bierbaum,$^1$ 
Karin A. Dahmen,$^2$ 
Carl P. Goodrich,$^3$
Julia R. Greer,$^4$
Lorien X. Hayden,$^1$ 
Jaron P. Kent-Dobias,$^1$ 
Edward D. Lee,$^1$ 
Danilo B. Liarte,$^1$ 
Xiaoyue Ni,$^4$
Katherine N. Quinn,$^1$ 
Archishman Raju,$^1$ 
D. Zeb Rocklin,$^1$ 
Ashivni Shekhawat,$^5$
and
Stefano Zapperi$^6$ 
\affil{$^1$Laboratory of Atomic and Solid State Physics, Cornell Univ., Ithaca, NY 14853-2501, USA; email:sethna@lassp.cornell.edu}
\affil{$^2$Physics Dept., Univ. of Illinois at Urbana-Champaign, Urbana, IL, USA, 61810}
\affil{$^3$School of Eng. and Applied Sciences, Harvard Univ., Cambridge, MA 02138}
\affil{$^4$Division of Engineering and Applied Sciences, California Institute of Technology, Pasadena, California 91125, USA}
\affil{$^3$Dept. of Physics, Univ. of Milano, Via Celoria 16, 20133 Milano, Italy}
\affil{$^3$Dept. of Physics, Univ. of Milano, Via Celoria 16, 20133 Milano, Italy}
\affil{$^5$Materials Sciences Division, Lawrence Berkeley National Laboratory, Berkeley, California 94720, USA}
\affil{$^6$Dept. of Physics
and Center for Complexity and Biosystems, Univ. of Milano, Via Celoria 16, 20133 Milano, Italy}
\vskip -0.3cm
}

\begin{abstract}
We give a bird's-eye view of the plastic deformation of crystals
aimed at the statistical physics community, and a broad
introduction into the statistical theories of forced rigid systems
aimed at the plasticity community. Memory effects in magnets, spin glasses, 
charge density waves, and dilute colloidal suspensions are 
discussed in relation to the onset of plastic yielding in crystals. 
Dislocation avalanches and complex dislocation tangles are discussed via
a brief introduction to the renormalization group and scaling. Analogies 
to emergent scale invariance in fracture, jamming, coarsening, and
a variety of depinning transitions are explored. Dislocation dynamics in
crystals challenges nonequilibrium statistical physics. Statistical physics
provides both cautionary tales of subtle memory effects in nonequilibrium
systems, and systematic tools designed to address complex scale-invariant
behavior on multiple length and time scales.
\end{abstract}

\maketitle

\tableofcontents

\section{Plastic deformation of materials: What is different?}
\label{sec:WhatIsDifferent}

Many of us, when trying to scoop ice cream, have induced {\em plastic deformation}: the spoon remains bent (\tBox{Spoon}).
Speculations about strategies for understanding our bent spoon 
represent the topic of this review.

\begin{jpsBox}{h}{Bent spoon and yield stress}{Spoon}

\begin{wrapfigure}[6]{L}{0.18\textwidth}
\vskip -0.3truein
\includegraphics[width=0.18\textwidth]{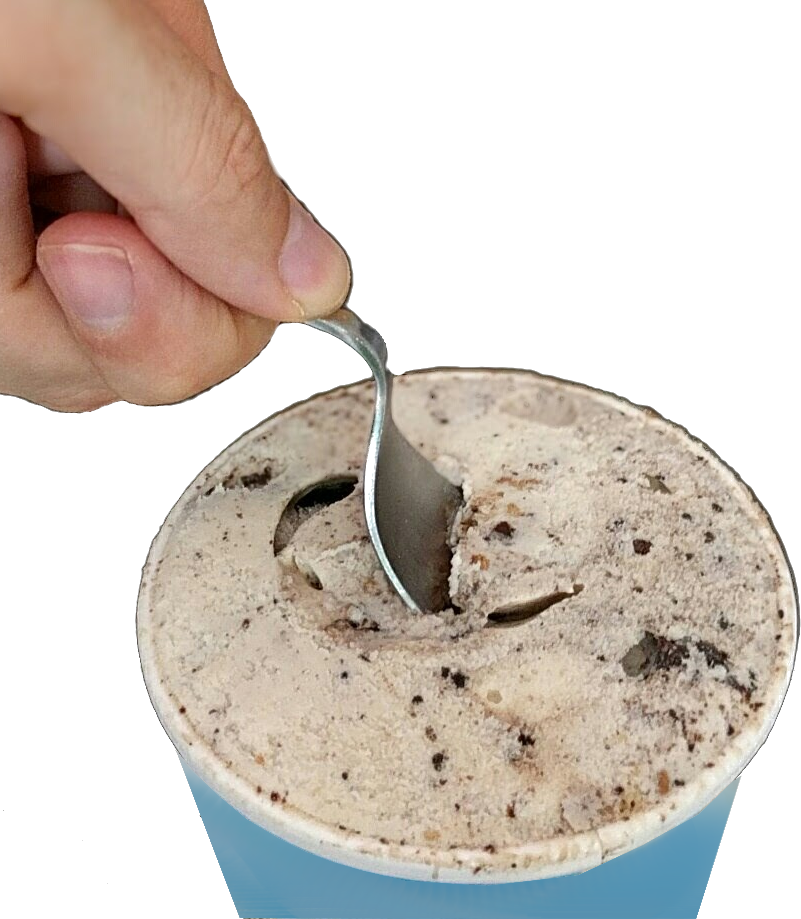}
\end{wrapfigure}

\noindent
A metal spoon will spring back into its original shape under ordinary use,
but in hard ice cream one may bend the spoon too far for it to recover. The spoon
is made up of many crystalline grains, each of which has a regular grid
of atoms. To permanently deform the spoon, atomic planes must slide past
one another.
This happens through the motion of dislocation lines. The dynamics,
interactions, and entanglement of these dislocation lines form the
microscopic underpinnings of crystal plasticity, inspiring this
review.

\end{jpsBox}

The physics and engineering communities have historically focused on
studying how plastic deformation
is {\em similar} to simpler systems.
Elastic materials respond to stress fields via strain fields; liquids respond
via viscous strain rates; complex fluids are often describable via
frequency-dependent viscoelastic responses. Many properties of crystals,
magnets, liquid crystals, superconductors, superfluids, and field theories
of the early universe can be described by focusing on long length scales
and assuming that the materials are locally close to equilibrium. 
Physicists use {\em generalized hydrodynamics}~\cite{Chandler,Forster,Martin}
or {\em Landau theory}~\cite{Landau}, and have systematized how conserved
quantities (particle number, momentum, energy) and broken symmetries
(magnetization, crystalline order) are assembled into {\em order parameter
fields}.%
\begin{marginnote}[]
\entry{Strain}{Fractional amount $\epsilon$ a crystal is stretched
or sheared, twisted or compressed.}
\entry{Stress}{Force per unit area $\sigma$ applied to a crystal. Both
$\epsilon$ and $\sigma$ are tensors, a fact we shall largely ignore.}
\end{marginnote}
Engineers apply {\em continuum field theories}~\cite{TadmorElliott}
using phenomenological material models to study materials behavior.
They incorporate {\em state variables} like dislocation
density, yield stress, or texture to describe the behavior 
of real materials at macroscopic scales. These models have been effectively
incorporated into the computational frameworks used to design everything from spoons to
airplanes. In {\bf Section~\ref{sec:PlasticityIntro}} we shall provide
a physicist's introduction to the challenges posed by plasticity in metals,
touching or ignoring many crucial features (slip systems, partial dislocations,
grain boundaries, {\em etc}),
but emphasizing how the broken translational symmetry
of crystals make them unusual among condensed matter systems.

\begin{marginnote}[]
\entry{Equilibrium}{In this paper, `equilibrium' and `nonequilibrium'
will ordinarily refer to {\em thermal} equilibrium (\tBox{Equilibrium});
most of our systems reach mechanical equilibrium -- a metastable state.}
\entry{Rigid}{Rigid systems have an energy cost to deforming some continuum
field -- magnetic, elastic, charge-density distortion, superfluid, \dots}
\end{marginnote}

Inspired by plasticity and failure of practical materials, the statistical
physics community has studied a remarkable variety of rigid nonequilibrium 
systems evolving under stress. They have focused on 
what makes magnets, sand, and other rigid systems under forcing
{\em different} from simpler systems~\cite{PWAIllCondensed}. 
This article summarizes results from many of these studies, from
magnetic hysteresis to jamming of granular materials, in brief vignettes
similar to \tBox{Spoon}.

There are striking
regularities that emerge in the deformation and failure of materials that 
lure us to search for a systematic theory. First, despite a bewildering
variety of material morphologies, most structural materials share certain
characteristic damage thresholds. Plastic deformation in practical materials
does not arise at
arbitrarily small applied stresses: the onset of deformation
characterizes the {\em yield strength} of materials (\tBox{WorkHardening}).
The yield stress is an approximation; creep and fatigue allow for hysteretic
changes below the yield stress. But by ignoring certain physical
mechanisms (e.g. things like vacancy diffusion and cross slip that 
typically go away at low temperatures)
we can study a theory that predicts a sharp threshold in quasi-static
deformations, and hence provides an
effective explanation of the observed, but less sharp, transitions in 
practical materials.

\begin{marginnote}[]
\entry{Burgers vector}{Topological charge of a dislocation: number of
extra planes of atoms crossed when encircling the
defect~\cite[Sect.~9.4]{Sethna06}.}
\entry{Creep}{Dislocation motions, like dislocation climb, that proceed
slowly and remain present below the traditional yield stress. They
are usually thermally activated. A component of rate-dependent plasticity.}
\end{marginnote}

The yield stress represents the division between elastic and plastic,
the division between reversible and hysteretic macroscopic
deformations. Fundamentally, equilibrium systems forget their
history: all other microscopic degrees of freedom are `slaves' to the
state variables. The chaotic local dynamics on the atomic scale scramble
the history of how the material was prepared; only those quantities
preserved by the dynamics can matter for the long-time macroscopic
behavior. In contrast, plastically deformed materials are not in local
equilibrium, and their properties depend on their history. The
blacksmith hammering the red-hot horseshoe and quenching it into water
is altering a complex microstructure which governs its toughness and
strength; a cast-iron horseshoe of identical chemical composition could
be brittle. In
{\bf Section~\ref{sec:Memory}} we shall briefly and broadly review how the
history of deformation has been reflected in other statistical mechanics
contexts:
can the memory of the material's past be effectively summarized in a
finite number of continuum variables? What can we glean from these other
systems about what might be special about the dislocation configurations
left behind after yielding and unloading in crystals?

Statistical physicists adore power laws and {\em emergent scale invariance};
when systems appear the same on different scales, we have powerful, systematic
tools to quantitatively predict the resulting behavior.
In {\bf Section~\ref{sec:Criticality}} we shall examine
evidence that dislocation evolution and plastic flow exhibits just this
kind of scale invariance ---
power-law distributions 
of dislocation avalanches and complex, perhaps scale invariant morphologies.
We shall discuss plasticity models which explain this as 
{\em self-organized criticality}, controlled by {\em work hardening},
and also discuss the possibility that these effects are due to a
proximity to a {\em critical failure stress}.
We shall discuss the theoretical {\em renormalization-group} framework
that has been used to analyze emergent scale invariance, introduce
related non-equilibrium statistical mechanics systems that have been studied
using these methods, and speculate about possible connections to plastic
deformation in crystals.  There we shall also discuss the tension between the 
{\em universality} traditionally expected at critical points and the 
strong non-universal dependence on material, morphology, and loading seen 
at the yielding transition.

\section{Plasticity: a physicist's introduction}
\label{sec:PlasticityIntro}

\begin{jpsBox}{th}{Crystalline rigidity}{CrystalRigidity}

\begin{wrapfigure}[6]{L}{0.65\textwidth}
\vspace{-0.5cm}
\includegraphics[width=\linewidth]{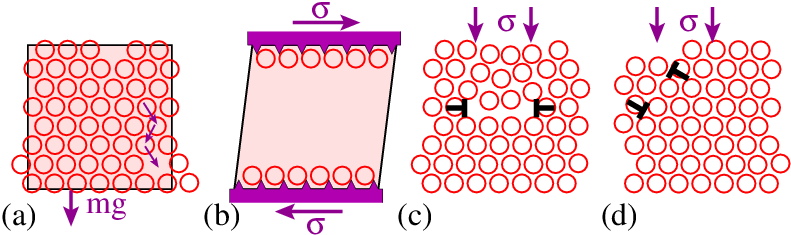}%
\end{wrapfigure}

\noindent
(a)~Under a force like gravity that couples to the mass density, an
equilibrium crystal {\em can} flow like a liquid (at a rate linear in
the gravitational force), via vacancy diffusion. (This is true at
temperatures above the {\em roughening 
transition}~\cite{RougheningDynamics}, where the equilibrium surfaces
all have steps that can absorb and emit vacancies.)
(b)~Crystals are rigid to
shears $\sigma$ that couple to the lattice ({\em i.e.}, the broken
translational symmetry).
(c)~A crystal sheared or compressed more than half a lattice constant will,
in equilibrium, lose a plane of atoms. Here a dislocation loop surrounds
a disk of missing atoms, which can grow only by {\em climb}, again via
thermally activated vacancy diffusion. (d)~The dominant shear relaxation
at low temperatures is {\em glide}, where a dislocation loop can grow along
a slip plane to relieve stress without involving vacancies (see
\tBox{Precursors}(a)). The nucleation rate for
the dislocation loops in~(c) and~(d) goes to zero faster than linearly as
the compression goes to zero; there is no linear viscous flow response
to forces that couple to the lattice. This distinguishes crystals from liquids.

\end{jpsBox}


\begin{marginnote}[]
\entry{Slip plane}{The plane accessible to a dislocation of a given Burgers
vector via glide.}
\entry{Slip systems}{The set of slip planes for the low-energy Burgers
vectors for a given crystal symmetry. The initial deformation of pure
metals usually `activates' only one or a few slip systems.}
\end{marginnote}

Almost any collection of atoms or molecules, when cooled, will form into
a rigid, solid object. This rigidity is an unappreciated, profound state
of matter.
Much is made of the precision measurements made possible by 
the Josephson effect in superconductors or the quantum Hall effect
in two-dimensional electron gases. But the rigidity of solids underlies
almost every measurement we do. For example, it allows
the mirrors of LIGO to be held multiple kilometers apart with an accuracy
much smaller than an atomic nucleus. From bridges to bones, the rigidity
of solids underpins our world.

Rigidity may seem straightforward. Atoms and molecules bond together,
and breaking bonds demands energy -- either thermal energy to melt the
material, or external forces to bend, shear, or fracture it. For glasses 
and amorphous materials, the key question is what makes them different
from liquids. Structurally similar to liquids, how and why do they get
trapped into a subset of the possible atomic configurations? What is the
nature of the glass transition, where the material stops flowing and 
develops an elastic rigidity to small external forces? We shall touch
upon current `jamming' theories regarding the rigidity of glasses in
\tBox{Jamming}.

Even the rigidity of crystals is subtle. Each atom in a crystal knows its
place: the regular array of atoms cannot flow.
A crystal has no linear shear-rate response to an external strain of the 
crystalline lattice (\tBox{CrystalRigidity}a, b). But it will shear under
high loads, plastically deforming through the motion of dislocations
(\tBox{CrystalRigidity}c, d).  What divides elastic from plastic?

How far can one strain a crystal before it will plastically deform -- 
rearranging its lattice to lower its energy? It is easy to see, but 
quite surprising, that the limiting stability region of an equilibrium
crystal is {\em microscopic}. Consider a crystal being 
compressed vertically (\tBox{CrystalRigidity}c,d), with free
boundary conditions along the horizontal. It is clear that the lowest
free energy configuration of the crystal will change -- removing one
horizontal plane of atoms, and moving them to the edge -- once the
compressive deformation reaches a lattice constant. Similarly, a
bent crystal can lower its energy by developing a low-angle grain
boundary once the net deformation is of order the lattice constant times
the logarithm of the number of atomic layers. This is quite different
from most broken symmetry systems; magnets, superfluids, and superconductors
can be twisted at their edges by large angles or phases before they
prefer to generate defects to relax. Crystalline rigidity is
{\em frail.}%
   \footnote{Smectic liquid crystals -- made of stacked two-dimensional
   liquid layers -- are also frail~\cite{LiarteBMKS16}.}


\begin{marginnote}[]
\entry{Peierls barrier}{The barrier per unit length to dislocation glide
due to the discreteness of the crystalline lattice; small for fcc metals.}
\entry{Kink}{The `soliton' mediating dislocation glide across the Peierls
barrier to the next lattice position. The barrier to kink motion is often
tiny ~\cite{VeggeSCJMR01}.}
\entry{Jog}{The soliton mediating dislocation climb, left behind, {\em e.g.},
when dislocations cross. Pins the dislocation.}
\entry{Stacking fault}{For simple fcc metals, a deviation from the fcc ABCABC
stacking -- typically low energy, since the nearest-neighbor structures
are unchanged.}
\entry{Partial dislocations}{The edge of a stacking fault, often
forming the core of a dislocation, described by fractions of a Burger's vector.}
\end{marginnote}

This frailty is not just a theoretical curiosity. Although perfect single
crystals (such as nanowhiskers) can support enormous strains before yielding,
high-quality nearly perfect fcc crystals have very low thresholds
for plastic deformation. Consider a dislocation line
segment in a crystal, pinned at two points a distance $L$ apart, perhaps
by inclusions, impurities, or other dislocations. (We ignore
effects due to the discreteness of the crystalline lattice.)
The component of the external stress $\sigma$ that couples to the dislocation
will cause it to bow out, forming a 
roughly circular arc, until its line tension and interaction energy balances
the external stress at a rough `curvature diameter' $D_c(\sigma)$. 
When $D_c < L$, the loop grows without bound, sometimes pinching off to form a
growing dislocation loop.

\begin{jpsBox}{h}{Dislocation pinning and work hardening}{WorkHardening}

\begin{wrapfigure}[10]{L}{0.7\textwidth}
\vskip -0.2truein
\includegraphics[width=0.7\textwidth]{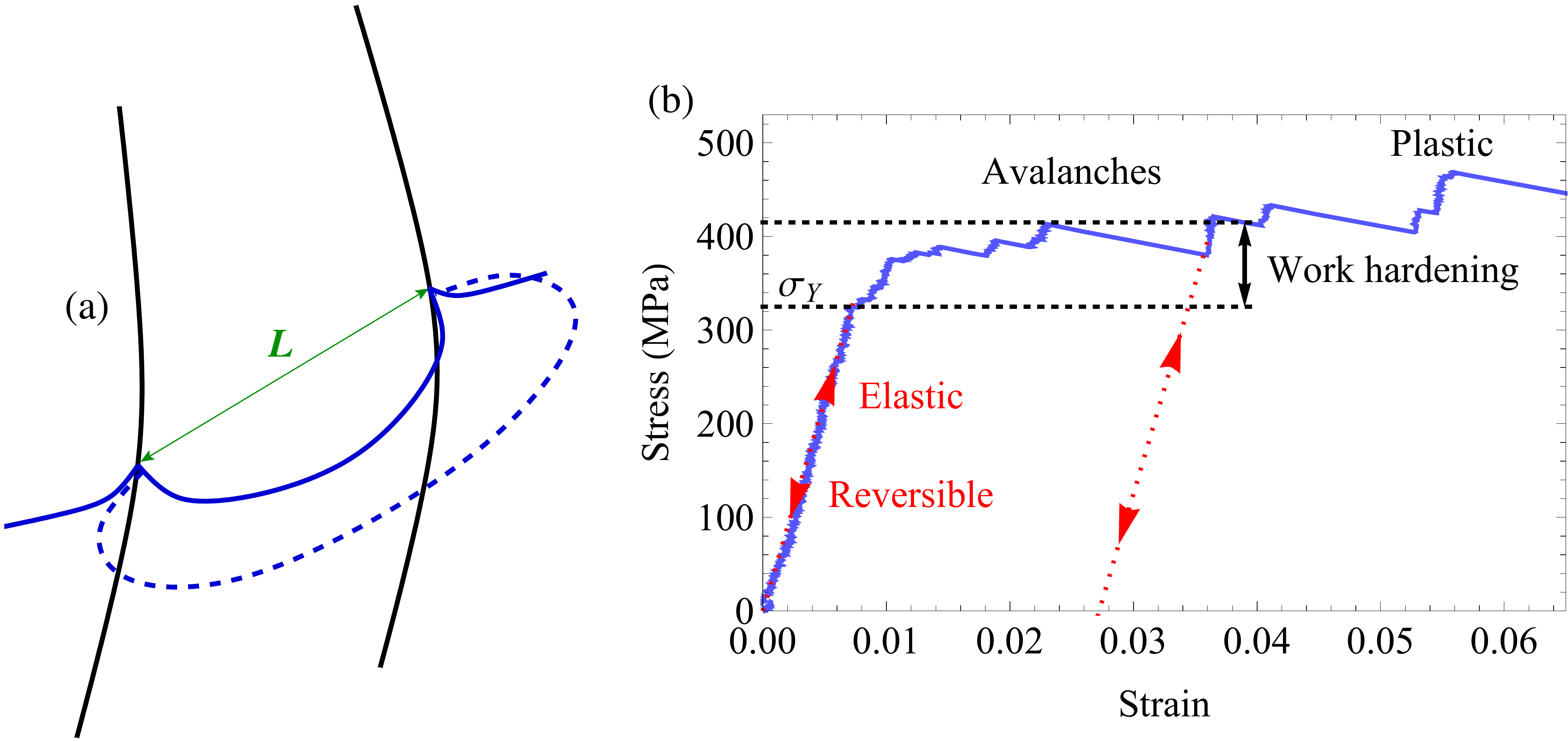} 
\end{wrapfigure}

\noindent (a) Dislocations get pinned on other dislocations, dirt,
inclusions, or their own jogs. Under stress, they will bow out
reversibly, with a `curvature diameter' which shrinks roughly linearly with
increasing stress, until it becomes less than the distance $L$ between pins.
At that point the dislocation balloons out and triggers an avalanche.
This argument explains the observed scaling of the yield stress 
$\sigma_Y\sim 1/L$ with the distance $L$ between pinning points
(see e.g. Ref.~\cite[Chapter~10]{hull11}). 
(b)~Micropillar deformation, showing avalanches~\cite{NiExpts}.
The response of a material is reversible
below a {\em yield stress} $\sigma_Y$, after which dislocation avalanches
lead to plastic deformation. When unloaded, the material will respond
reversibly until reloaded to roughly the previous maximum stress. Subsequent
plastic flow increases the length and density of dislocations, leading to
more pinning and {\em work hardening}.

\end{jpsBox}

The durable crystals familiar to us have rather large densities of pinning
sites, often arising from the tangle of dislocations formed by previous 
plastic deformation. There is an energy barrier impeding
the crossing of two (or more~\cite{BulatovMultiJunction}) 
dislocations;%
   \footnote{There is no {\em topological} reason to prevent dislocations
   from crossing~\cite[Chapter 9]{Sethna06}, although often a jog 
   will be left behind, which then can impede glide.}
they often instead merge into dislocation {\em junctions}, which act as 
pinning points. If the density of dislocation lines crossing a unit
area is $\rho$, a dislocation passing through the tangle will typically
have pinning intersection points separated by $L\sim 1/\sqrt{\rho}$
(\tBox{WorkHardening}(a)). This implies a yield stress
$\sigma_Y \propto \sqrt{\rho}$ --- the Taylor relation~\cite{Taylor34}.%
   \footnote{One must note that grain boundaries can also act as pinning
   sites for dislocations, but the yield stress for grains of size $L$
   does not scale as $1/L$, but approximately as $1/\sqrt{L}$ (the Hall-Petch
   relation), due to the need for several dislocations
   to cooperate in punching the lead dislocation through the wall (e.g.,
   in a `dislocation pileup'). This different scaling will be a challenge
   to scaling theories like those discussed in
   {\bf Section~\ref{sec:Criticality}}.}
This power-law scaling relation, while simply derived, foreshadows the 
types of predictions derived from emergent scale invariance in
{\bf Section~\ref{sec:Criticality}}.

As plastic deformation proceeds, the dislocation lines stretch and multiply,
increasing their density and further increasing the yield stress. This {\em
work hardening} makes plastic deformation self-limiting in crystals; a weak
spot becomes stronger when it yields.%
   \footnote{Metallic glasses, on the other hand, become weaker as they
   shear, leading to failure via slip bands~\cite{budrikis13,sandfield15}.}
Hence a pristine copper wire (or an ordinary metal clothes hanger) can be bent
quite easily into a tight curve. But once bent, it becomes far more resistant
to further deformation; bending takes far less force than unbending.

\begin{marginnote}[]
\entry{Screw dislocation}{A dislocation directed parallel to its Burgers
vector, forming a `spiral stair'. The screw portion of a dislocation
loop can glide in any direction.}
\entry{Cross slip}{The motion of the screw portion of a dislocation loop
into a different glide plane, often involving thermally activated 
restructuring of partial dislocations.}
\end{marginnote}

Plastic deformation is the study of non-equilibrium 
collective dynamics of spatially extended topological defect lines, with
long-range elastic interactions and complex constraints on the defect motions.
Each of these features has separately become a focus of statistical physics
in the past decades. Indeed, many of us have been inspired to study 
memory effects and critical phenomena in these other systems because of 
seeming relations to the challenging practical problems illustrated
by bent spoons. Let us now proceed with a distillation of some of the key
ideas developed in statistical physics that could be useful or inspirational
for the study of plastic deformation of crystals.

\section{Memory and state variables}
\label{sec:Memory}

It is a truism in metallurgy that the mechanical properties of crystals
depend on their thermal and deformation history -- the `heating and beating'
suffered by the material in the past~\cite{dieter1986mechanical}.
Mechanical plastic deformation of a crystal
typically increases the threshold for further deformation 
(work hardening); heating it to high temperatures anneals the crystal
back to a plastically soft state. As discussed in
{\bf Section~\ref{sec:PlasticityIntro}},
different theories of plasticity attempt to encapsulate this history 
dependence into a variety of state variables. 
The simple sample yield stress we have discussed, for example,
could be promoted into a spatially-dependent yield
stress of an inhomogeneously processed specimen, or to an entire yield surface
in the six-dimensional space of stresses, or to a scalar, tensor, or 
slip-system specific dislocation density, {\em etc.}

The history-dependence of other rigid systems has been an active field 
in statistical mechanics. Some have focused on how non-equilibrium systems 
might be similar to equilibrium phases, with a thermodynamics resulting from
a guiding principle similar to maximizing entropy.
Some have explored the weird ways that particular rigid systems respond
to external forcing.
And some have focused
on how non-equilibrium systems may be governed by critical points
({\bf Section~\ref{sec:Criticality}}), with 
scaling behavior similar to continuous equilibrium phase transitions. 

\begin{jpsBox}{h}{Equilibrium statistical mechanics}{Equilibrium}

\noindent 
Equilibrium systems are simple~\cite[Chapter 4]{Sethna06} first because
their dynamics is chaotic. Chaotic systems rapidly lose all information about
their previous state, except for conserved quantities 
(energy, volume, number of particles, \dots), and fields associated to
broken symmetries
(magnetization, crystalline strain, superfluid phase, \dots). Their
time-average behavior then becomes a weighted average over their `attractor'.
Thermal equilibrium systems are simple second because their
energy-conserving Hamiltonian dynamics
preserves volume in phase space (Liouville's theorem), telling us that
the attractor includes all possible states consistent with the
preserved information, weighted by phase space volume (a `maximum entropy'
state). This leads directly to free energies and
thermodynamics. Temperature, pressure, chemical potential, and stress
arise as Lagrange multipliers to constrain the conserved energy, volume,
number, and strain.

\end{jpsBox}

The plasticity of crystals forms the prototypical example of a nonequilibrium, 
history-dependent rigid system. It has inspired and guided careful work
on memory effects in other statistical mechanical contexts. We may hope that
some of the unusual memory ramifications observed in these simpler systems
might be relevant for plasticity.
In this section, we briefly highlight studies of the first two types
(nonequilibrium thermodynamics, {\em vs.} weird memory effects), and
discuss how they might give insight into practical plasticity problems. 
We will turn to consider nonequilibrium critical points in
{\bf Section~\ref{sec:Criticality}}.

\begin{jpsBox}{h}{Compactivity {\it vs.} angoricity}{CompactivityAngoricity}

\renewcommand{\arraystretch}{1.7}
\begin{wrapfigure}[11]{L}{0.8\textwidth}
\vspace{-0.25cm}
\begin{tabular}{| l | c | c | c |}
\hline
            				& Boltzmann						& Edwards						& HC-EB
\\ \hline \hline
Conserved quantity	& Energy, $E$					& Volume, $V$					& Force-moment, $\hat{\Sigma}$
\\ \hline
Entropy				& $S =k_B \ln \Omega_B (E) $	& $S =\lambda \ln \Omega (V)$	& $S =\lambda \ln \Omega (\hat{\Sigma})$
\\ \hline
Intensive quantity	& Temperature					& Compactivity					& Angoricity
\\ \hline
					& $\displaystyle\frac{1}{T} = \frac{\partial S(E)}{\partial E}$ 
					& $\displaystyle\frac{1}{X} = \frac{\partial S(V)}{\partial V}$
					& $\alpha_{\mu \lambda} = \displaystyle\frac{\partial S(\Sigma_{\mu \lambda})}{\partial \Sigma_{\mu \lambda}}$
\\ \hline
Distribution			& $\exp [-E / (k_B T) ]$			& $\exp (-V/X )$					& $\exp (-\alpha_{\mu \lambda} \cdot \Sigma_{\mu \lambda} )$
\\ \hline
\end{tabular}
\end{wrapfigure}

\noindent
There has been an ambitious attempt to use statistical mechanics
to derive a thermodynamics of dry grains and dense non-Brownian
suspensions~\cite{bi15}. Grains of sand do not move unless pushed;
energy and temperature are unimportant. 
Edwards and Oakeshott~\cite{edwards89} proposed a `microcanonical'
ensemble for powders that maximized an entropy with 
volume as a conserved state variable, with the distinction from phase
space that the states $\Omega(V)$ were restricted to jammed configurations
(see \tBox{Jamming})~\cite{martiniani16}. Recent experiments suggest that the zeroth law
of thermodynamics does not hold for the resulting `compactivity' temperature.
By incorporating force and torque balances,
Henkes \& Chakraborty~\cite{henkes05} and
Edwards \& Blumenfeld~\cite{edwards05,blumenfeld09}
derive a thermodynamics where 
a force-moment tensor $\hat{\Sigma}$ plays
the role of energy, and the intensive quantity $\hat{\alpha}$, named
\emph{angoricity} (`stress' in Greek), plays the role of
temperature. The zeroth law so far appears to hold for angoricity. 
\end{jpsBox}

Sam Edwards~\cite{edwards89} proposed the most influential recent
analogy between nonequilibrium systems and equilibrium statistical
mechanics~(\tBox{Equilibrium}): a thermodynamic theory of packed
granular powders.%
    \footnote{Note that low-density shaken granular materials {\em fluidize};
    such systems do explore their available states and in many regimes
    can be approximated well by theories drawing from
    thermodynamics~\cite{jenkins2016kinetic}.}
He posited a `phase space' of force-balanced jammed arrangements of grains
producing a kind of granular entropy, with volume replacing energy
as the conserved quantity and temperature replaced by 
a {\em compactivity} field. Recent experiments~\cite[see \tBox{CompactivityAngoricity}]{henkes05,bi15}
show very generally that such a 
description does not satisfy the zeroth law (two kinds of particles in
equilibrium with a third must be in equilibrium with one another).
Indeed, it is hard to see how the admittedly complicated
dynamics of grain motion as it is tamped or sheared would rearrange particles
to transmit extra volume effectively through the system.
However, the forces between touching
particles can rearrange dramatically under changing loads even for fixed
particle contacts. (Puckett and Daniels~\cite{PuckettD13} mention that 
forces and torques balance at each contact, while volume is merely
conserved globally.)
This perhaps could form the basis of a thermodynamic theory
(\tBox{CompactivityAngoricity}). 

There are serious challenges to similar analogies between 
dislocations and equilibrium statistical mechanics. 
Dislocation {\em energies} are strongly coupled to their atomic
environments; they lose heat as they nucleate, move, and tangle, and hence
it would seem that no direct `dislocation temperature' should
exist~\cite{LangerBL10} distinct from the thermodynamic
temperature. Many of these same concerns, however, apply to granular
systems~\cite{MakseKurchan,AbateD08}, glassy simulations~\cite{BerthierB02}
and foams~\cite{OnoODLLN02}, all of which have shown evidence of
effective temperatures and maximization of entropy~\cite{martiniani16}.
Dislocations also exchange force with the crystal lattice;
(pinned to inclusions, jogs, `sessile' dislocation junctions, \dots),
presumably also arguing against a `dislocation angoricity'
(\tBox{CompactivityAngoricity}). 

\begin{jpsBox}{h}{Magnetic memory of heating and beating}{WeirdMemory}

\begin{minipage}{1.1\textwidth}
\centering
\includegraphics[width=\textwidth]{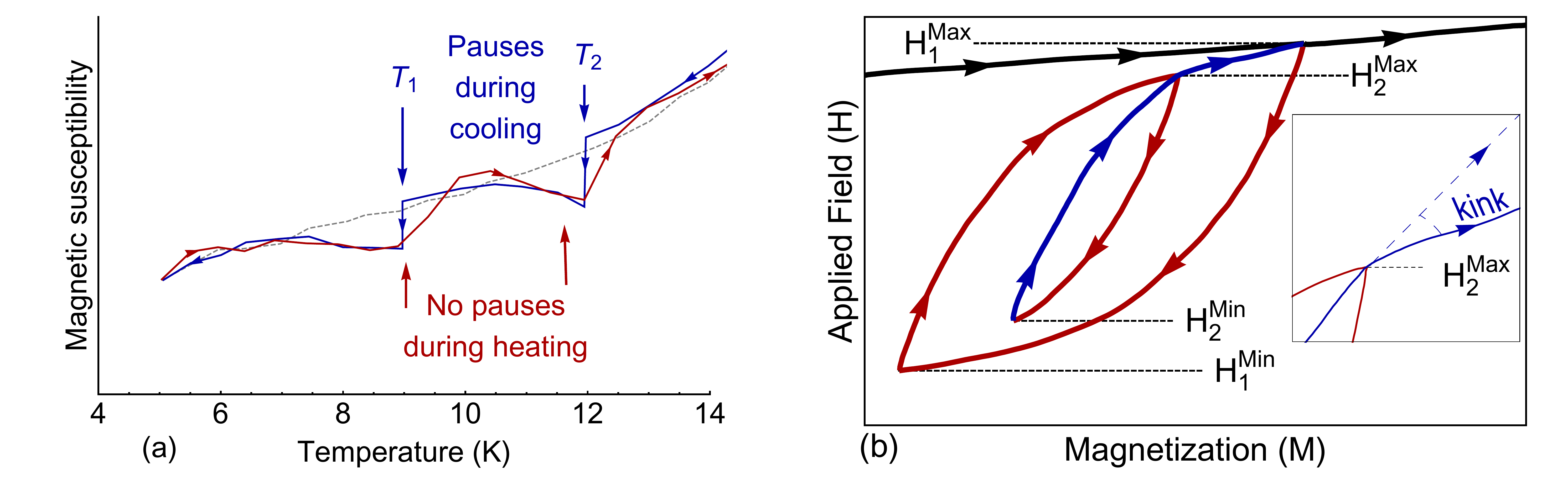}
\end{minipage}
\noindent
(a)~Spin glasses are frustrated magnetic systems, which have modes that
are active in narrow ranges of temperatures. Annealing on cooling at 
temperatures $T_1, T_2$ leaves fewer susceptible modes upon heating through
the same temperatures~\cite{jonason1998memory}.
(b)~Disordered ferromagnets at zero temperature have hysteresis
loops that return to precisely the same microstates after the external field
is decreased from $H^{\mathrm{max}_1}$  and
returned to the previous maximum $H^{\mathrm{max}}$, even though
the avalanches differ on raising and lowering the field.
This is a {\em return-point memory} of the previous external field
history~\cite{SethnaDKKRS93}. Each return to a previous maximum
results in a kink in the $M(H)$ curve.

\end{jpsBox}

More prevalent in nonequilibrium statistical mechanics are studies of how
particular rigid systems behave weirdly under forcing.%
  \footnote{A quantitative `dislocation thermodynamics' would seem incompatible
  with such weird behavior.}
These other systems have attracted interest not 
because of the behavior as they are unloaded and reloaded, but rather
because of the character of the microstates selected by the
cycle of unloading and reloading.
What can we glean from these about
crystal plasticity?

\begin{marginnote}[]
\entry{Rate-independent plasticity}{Deformations that happen slowly
compared to dislocation glide process, but fast compared to creep; the
focus of this discussion.}
\end{marginnote}
 
The textbook picture of yield stress in crystals~\cite{hull11,dieter1986mechanical}
is particularly simple.
Raising the stress to the yield stress and back to zero supposedly leaves 
the system 
in precisely the same state (ignoring creep); raising the stress beyond
the yield stress leads to rearrangements of dislocations into 
new metastable configurations. Dislocations move
and tangle under increasing stress, like snow gets pushed when a snow plow
moves forward on the street. Under decreasing forcing, the dislocation
tangles presumably remain in place just as does the snow.

Magnets and other hysteretic systems have hysteresis upon decreasing and 
increasing the external field, but certain magnetic systems
do return to precisely the same state upon reloading (the 
{\em return-point memory effect}, \tBox{WeirdMemory}b).
The avalanches they undergo upon unloading differ from those upon reloading.
Upon increasing the external field $H$, $M(H)$ will have a kink 
at the point when it `rejoins' the old trajectory. For a system
trained into multiple hysteresis subloops, a kink arises
upon raising the field above each (monotonically increasing) historical peak
in the field history. In principle, a magnet could encode an indefinite number
of such kinks,
suggesting that no finite number of state variables can perfectly
encode the behavior. Memory effects also arise from the thermal history in
spin glasses, which encode pauses in the thermal history upon cooling
(\tBox{WeirdMemory}a).

\begin{jpsBox}{h}{Phase organization}{PhaseOrganization}

\begin{minipage}{1.1\textwidth}
\centering
\includegraphics[width=\textwidth]{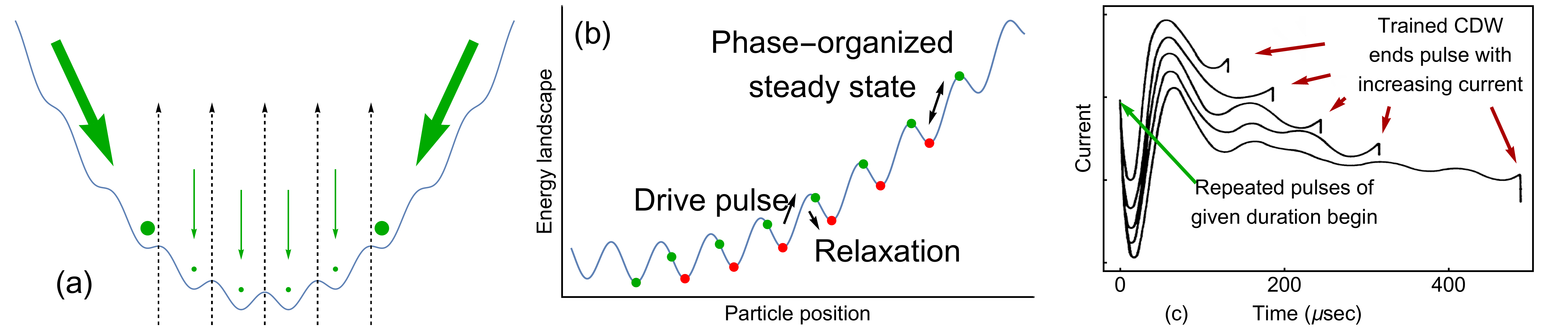}
\end{minipage}

\noindent
Does a nonequilibrium system dynamically dumped into a stable state
select typical or unusual metastable states?
(a)~Consider a particle in a periodic potential,
connected by a spring to a `nail' at the origin~\cite{coppersmith1987simple}.
A typical initial configuration will slide down the potential until the
first local minimum, which will usually be on the edge of the range of
local minima -- just barely stable to inward forces. A collection of such
nails (representing many local regions of a material), upon random 
initialization, will typically have most nails on their edges -- a marginally
stable hyper-corner of the hyper-cube of possible system
configurations. (b)~For an ensemble of nailed particles trained by a pulse
of a given duration~\cite{coppersmith1987pulse}, the pulses drive the particles
{\em up} the spring potential, until a phase-organized `edge' state 
in a marginally stable configuration at the {\em end} of the pulse.
(c)~In charge-density-wave materials, this leads to a current that always
{\em rises} at the end of the
pulse~\cite{coppersmith1987pulse,tang1987phase}.

\end{jpsBox}

Another rigid state of matter is the {\em charge density wave} -- a 
modulation in the electron density of a crystal whose wavelength can be
unrelated (incommensurate) to the crystal lattice, and which is often
pinned by impurities.
The charge density wave in some materials can slide under a large
enough external voltage. When slid repeatedly for
a fixed pulse time $t$, they settle into a stable limit cycle.
Unlike the simple
model of plasticity and the return-point memory magnets, here the 
hysteretic behavior must be {\em trained} by a few cycles. These
systems exhibit a striking collective effect, termed 
{\em phase organization}~\cite{coppersmith1987pulse,tang1987phase,coppersmith1987simple}; the current at the end
of the training pulse always is increasing -- reflecting the marginal
stability of the `first' periodic limit cycle found by local regions
in the material. In the rough hypercube of possible periodic states, the 
ones selected are at hypercorners (\tBox{PhaseOrganization}). 
This example warns us that the `stuck' states actually
occupied in non-equilibrium systems can be distinct from typical metastable 
states in striking ways. 


\begin{jpsBox}{h}{Avalanche precursors}{Precursors}


\begin{wrapfigure}[8]{L}{0.7\textwidth}
\vspace{-0.5cm}
\includegraphics[width=0.7\textwidth]{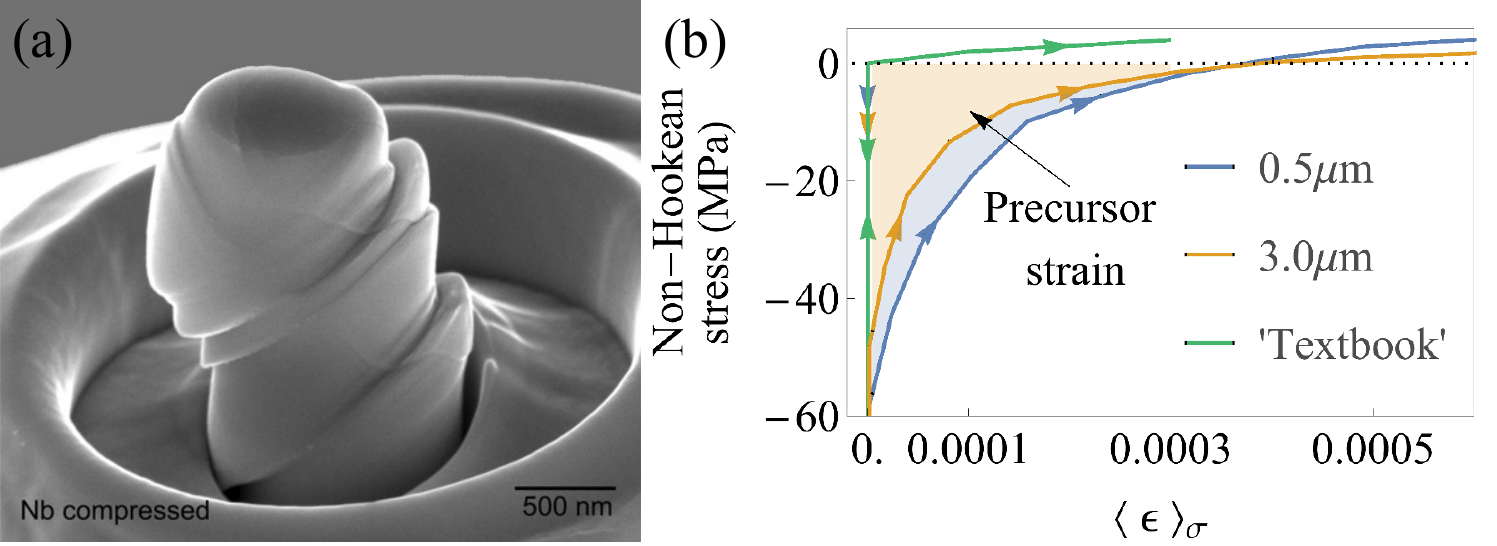}
\end{wrapfigure}

\noindent
(a)~{\bf Micropillars under compression} (when oriented properly) yield along a single glide plane. (b)~{\bf Dislocation avalanches} become visible in these small-scale crystals. The macroscopic theory in textbooks~\cite[Ch. 26]{AsaroLubarda} predicts that under deformation the yield stress `self-organizes' to the current stress $\sigma_\mathrm{max}$, and the material obeys Hooke's law as $\sigma_{ij} = C_{ijkl} \epsilon_{kl}$ upon unloading and reloading up to $\sigma_\mathrm{max}$. Here we show preliminary precision measurements by Ni and Greer~\cite{NiExpts} of the stress versus average strain upon reloading, for two different copper micropillars. The individual experiments clearly show precursor avalanches upon reloading, which are not a part of the macroscopic theory~\cite{silva13,nair15,philpot12}. These precursors average together into the stress-strain curves shown. The smaller net precursor strain occurs in the larger pillar; it is possible that the precursor avalanches entirely disappear in macroscopic samples.



\end{jpsBox}

We should note that the simple model of a yield stress, separating
purely elastic behavior from irreversible deformation, and determined
purely from the previous stress maximum, perhaps is oversimplified;
indeed, small amounts of hysteresis and training are probably to be
expected in general.
For example, if the
stress swings {\em negative}, the dislocations will
start to rearrange at absolute stresses somewhat lower than the yield
stress (the Bauschinger effect~\cite{dieter1986mechanical}, Ch.~6) -- reverse avalanches certainly happen
as the stress direction is reversed.
Since there are local stresses in a tangle due
to the dislocation interactions that are comparable to the yield stress,
it should be true that there are occasional
`backward' avalanches on unloading: the local material does not know when
the point of zero external stress is crossed. Similarly there likely will be
some precursor avalanches upon reloading below the yield stress. 
`Trained' reversible states may have avalanches that differ upon
loading and unloading but whose effects cancel. 
\tBox{Precursors} discusses preliminary results of micropillar
compression experiments by Ni and Greer, with no avalanches upon
unloading, but precursor avalanches upon reloading before reaching
the previous stress maximum. (Note that 
these precursor avalanches could be a finite-size effect, with smaller
plastic strain for larger pillars.)


\begin{jpsBox}{h}{Sloppy models}{SloppyModels}


\setlength{\columnsep}{0pt}%
\begin{wrapfigure}[7]{l}[0pt]{0.75\textwidth}
\vskip -0.25truein
\null\hskip -0.1truein 
\includegraphics[width=0.75\textwidth]{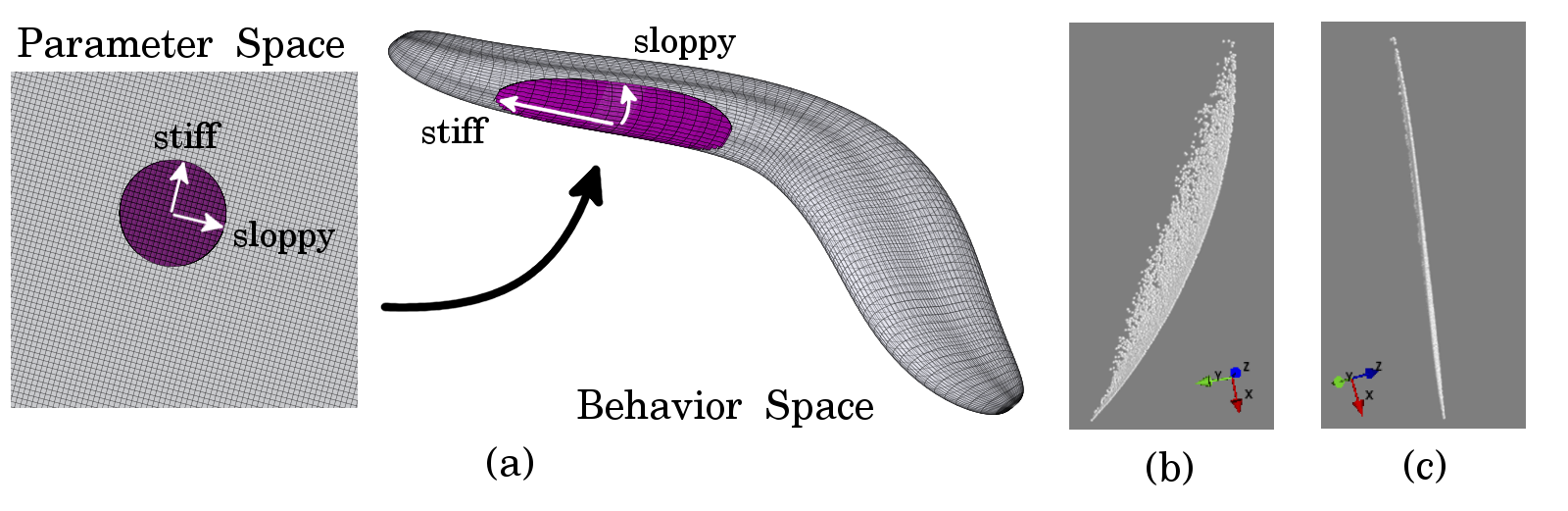}
\end{wrapfigure}

\noindent 
Science is possible because of a kind of {\em parameter compression}~\cite{MachtaCTS13}. The behavior of complicated systems (cell biochemistry, interatomic potentials, insect flight, critical phenomena) can be expressed in a relatively simple way, controlled by a few {\em stiff} combinations of the microscopic parameters. (a) {\em Information geometry} describes this in terms of the manifold of all possible macroscale behaviors, which often forms a hyperribbon~\cite{TranstrumMBDMS15}; `sloppy' parameter combinations move along the thin directions of this model manifold. (b,c) The model manifold for  fitting curves to radioactive decay forms just such a hyperribbon (Fig. from~\cite{TranstrumMBDMS15}). Hence, it is challenging to extract lifetimes from a sum of exponential decays. Similar decay rates and amplitudes can be `traded' for one another along {\em sloppy} directions, with only \textit{stiff} combinations constrained to keep the sum fixed. If we view the temperature and strain history in crystal plasticity as parameters, and the macroscale anisotropic strength and toughness as behavior, this parameter compression could explain the emergence of effective plasticity theories with relatively few state variables encapsulating the `stiff' microscale combinations.

\end{jpsBox}

What do these examples suggest about state variables in theories of 
plastic flow in materials? 
It seems likely that crystalline
dislocation networks do store their history in a way similar to that of
spin glasses, magnets, and sliding charge-density waves 
(\tBox{WeirdMemory}).
One might naively predict that deforming 
a crystal in some pattern as it is cooled could, upon reheating, lead 
to a reversed `echo' deformation.
In fact, bent crystals do not return to their 
original shapes when they are thermally annealed: a squashed nanopillar
(\tBox{Precursors}a) does not stand up again upon heating.
Instead, the deformation history after thermal annealing is 
reflected in surface steps (\tBox{Precursors}a) and (for polycrystals)
in the grain orientation distribution
(called {\em texture}~\cite{kocks2000texture,randle2000texture}).
We must note, though, that {\em Shape-memory alloys} do regain their
original shapes upon reheating, due to the existence of a unique
austenitic high-temperature state~\cite{bhattacharya03}.

The fact that many historical parameters can be read from
the current state does not preclude a useful low-dimensional description.
Just because the current state of a material can encode its whole history does
not mean details of the history affect the behavior in {\em significant}
ways. For example, while a magnet after `ring-down'
(\tBox{WeirdMemory}) will show a cusp in the magnetization 
curve at every historical ring-down maximum, the cusps rapidly become tiny; 
the details of subloops of subloops of hysteresis loops are not crucial
for understanding magnets. Indeed, this is a broad feature of nonlinear
systems and models depending on many parameters; their behavior is usually well
described by a few `stiff' combinations of parameters (see
\tBox{SloppyModels}). Perhaps yield stress, damage, porosity and other
local variables
are capturing the stiff microscopic combinations governing macroscale behavior.

\begin{jpsBox}{h}{Hysteresis, reversibility, and irreversibility}{HITRIT}

\begin{wrapfigure}[7]{r}[0pt]{0.6\textwidth}
\label{fig:couette}
\vskip -0.3truein
\null\hskip -0.4truein \includegraphics[width=0.6\textwidth]{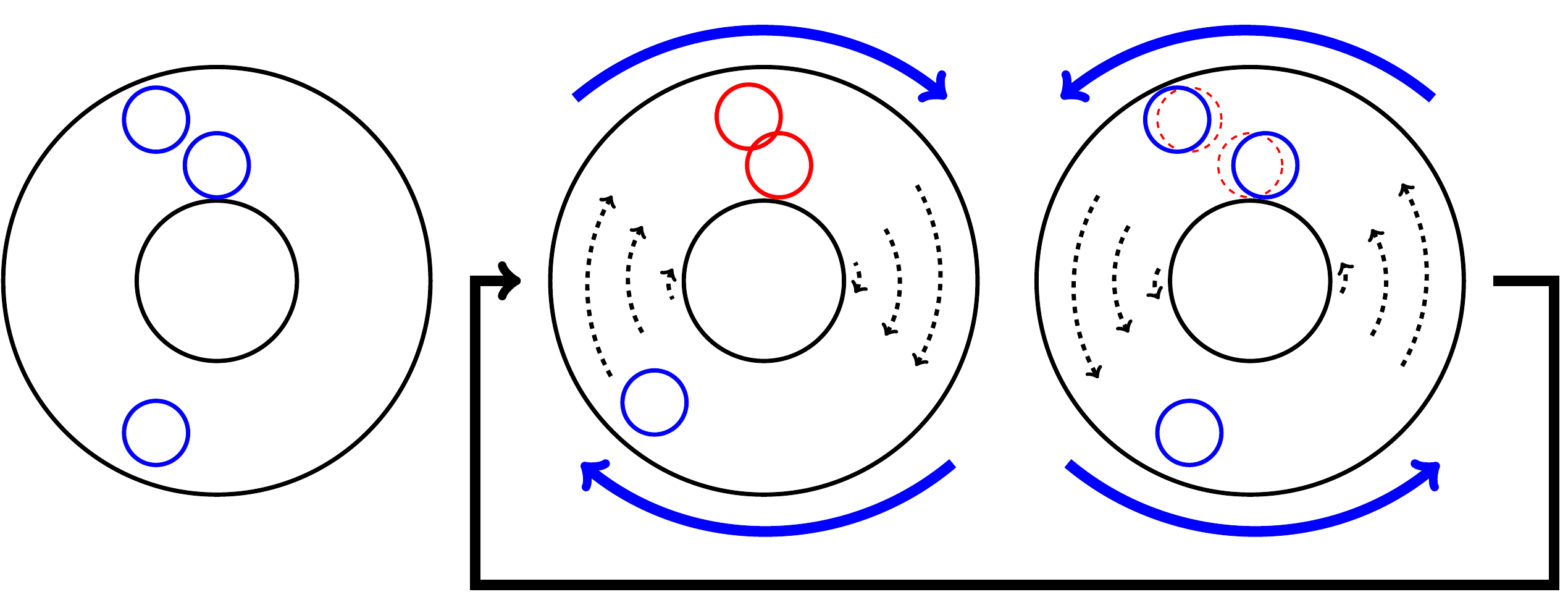}
\end{wrapfigure}
\noindent
Forced viscous fluids (at low Reynold's numbers) reverse their motion when the forcing is reversed. When tiny 
colloidal particles are added, their trajectories reverse if they do not collide during the forward motion.
Repeated oscillations at fixed amplitude can train the particles not to collide, mimicking crystals forced below
their yield stress~\cite{Corte08,Pine05,Paulsen14,Reichhardt09,Keim13,Keim14,Menon09}.
Increasing the amplitude past the previous maximum training point yields new colloidal collisions.
Similar transitions (except with hysteresis in the trained state) have been seen in
amorphous solids~\cite{Regev15,Regev13,Fiocco13,Jeanneret14,Nagamanasa14,Rogers14}, 
granular systems \cite{Mobius14,Schreck13,Slotterback12,Royer15}, 
dislocations \cite{Zhou14}, 
and super-conducting vortices \cite{Okuma11, Mangan08, Perez11,Lopez99,Miguel03,Shaw12,Okuma12}.
If the density of colloidal particles is high, or the shear amplitude
is large, each collision can trigger on average one or more collisions in the next oscillation, leading to a 
Reversible--to--Irreversible Transition (RIT), also observed in the other
systems listed above; the necessary cycling time, and the length scales of
the correlated rearrangements, both diverge with power laws
as discussed in {\bf Section~\ref{sec:Criticality}}.

\end{jpsBox}

Finally, we must mention recent work on Reversible-to-Irreversible Transitions
(RITs, \tBox{HITRIT}).%
   \footnote{Here we must distinguish `local irreversibility' from
   rearrangements like avalanches which do not reverse along the same 
   path upon unloading,
   and `global irreversibility' when the loading gets large enough that 
   cycling never settles down. The RIT in most cases is a transition from
   the former to the latter.}
These are systems 
that can be `trained' into periodicity under low amplitude cycles.
In many systems, the number of cycles to train the system can grow 
to infinity at a critical
amplitude, after which irreversible changes continue forever.
This is our
first example of a critical point exhibiting power laws and scaling functions
-- the topic of {\bf Section~\ref{sec:Criticality}}.

\section{Emergent scale invariance in plastically deformed crystals}

\label{sec:Criticality}

\begin{jpsBox}{h}{Renormalization group}{RG}

\noindent

\begin{wrapfigure}{L}{0.3\textwidth}
\vskip -0.4cm
  \centering
  \includegraphics[width=0.9\linewidth]{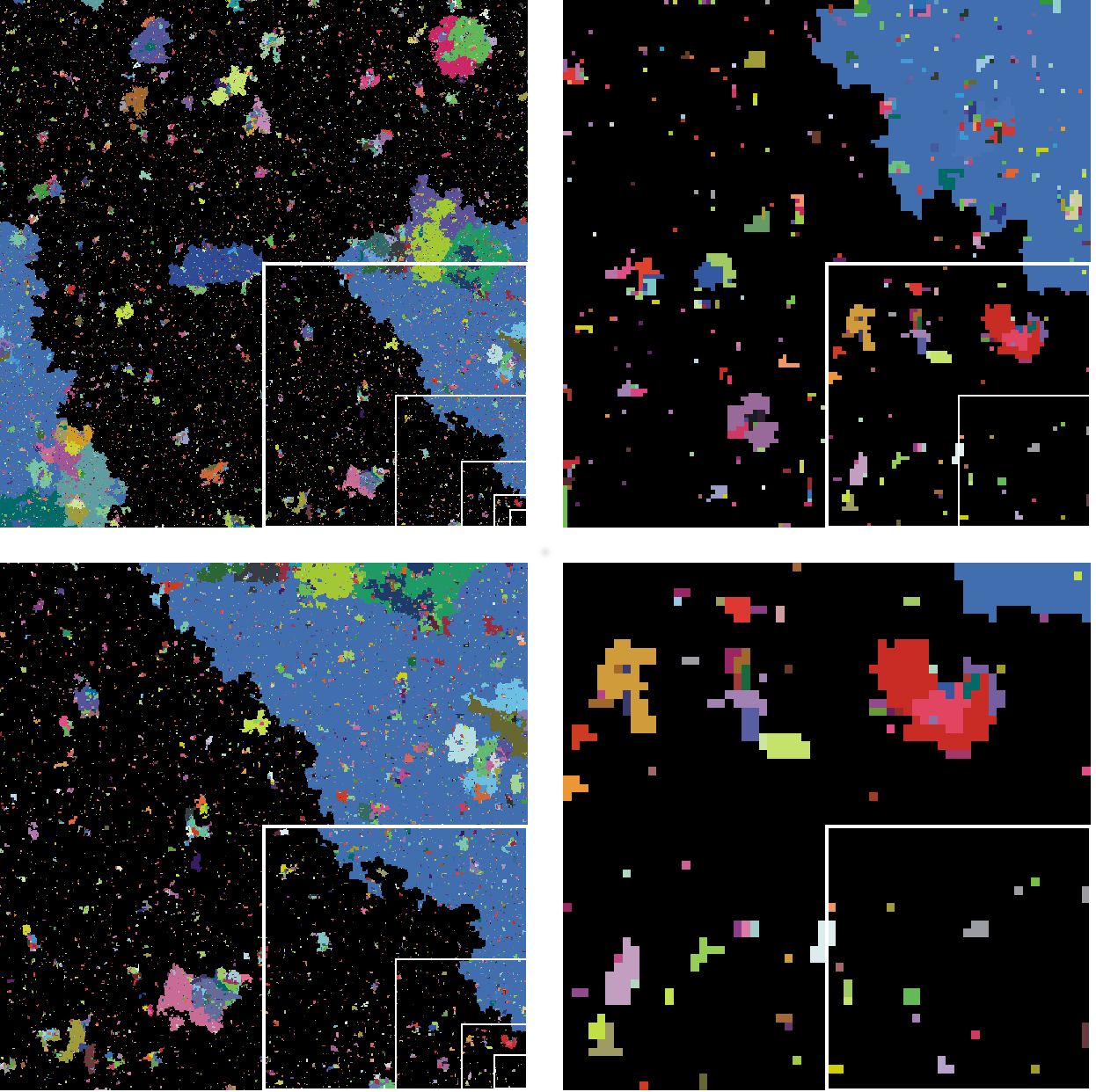}
 (a) \\
 \includegraphics[width=0.9\linewidth]{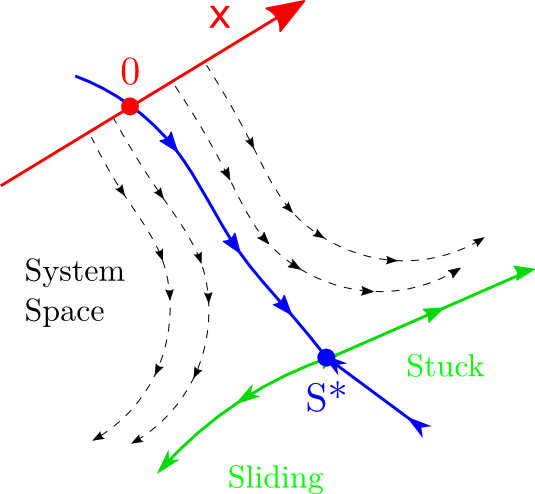}
 (b)
\end{wrapfigure}

\noindent
The renormalization group (RG)~\cite{goldenfeld1992lectures} is an amazing abstraction which works in
a `system space', with dimensions parameterizing both different possible
statistical models and real experiments. The RG studies how the rules governing
a system change with length scale. By coarse-graining (ignoring microscopic
degrees of freedom up to a scale $b$), one derives rules with {\em renormalized}
parameters -- a dynamical flow in system space (arrows in b).
(a)~A RG fixed point $S^*$ exhibits {\em scale invariance}, so for example 
the avalanches in space shows a statistical self-similarity upon zooming in 
to the lower right-hand corner.~\cite{SethnaDM01}

(b)~Tuning a parameter moves along a line (red) in system space; if it crosses
the `stable manifold' of $S^*$ (green) the system will be self-similar
after coarse-graining to long length scales.

Consider some observable $Z$ (say the avalanche size distribution) as a
function of parameters $x$, and $y$, (which could be avalanche size, system
size, stress, or the work-hardening coefficient). If $Z$, $x$ and $y$ are 
eigenvectors of the linearized flow around $S^*$,
then under coarse-graining
$Z(x,y) = b^{-\lambda_z} Z (x b^{\lambda_x}, y b^{\lambda_y})$, where $\lambda$
is the corresponding eigenvalue.
If we coarse grain until $x$ flows to one (so $x b^{\lambda_x} = 1$), then
\begin{equation*}
 Z = x^{\frac{\lambda_z}{\lambda_x}} Z(1, y x^{-\frac{\lambda_y}{\lambda_x}}) = x^{\frac{\lambda_z}{\lambda_x}} \mathcal{Z}(y x^{-\frac{\lambda_y}{\lambda_x}}), ~~ \mathrm{or~alternatively} ~~
 Z = y^{\frac{\lambda_z}{\lambda_y}} \tilde{\mathcal{Z}} (y x^{-\frac{\lambda_y}{\lambda_x}})
\end{equation*}
The observable is a power law times a \textit{universal scaling
function} $\tilde{\mathcal{Z}}$ of an invariant combination of
parameters $y x^{-{\lambda_y}/{\lambda_x}}$. In plasticity, the 
probability $P \equiv Z$ of an avalanche of size $S\equiv y$ with a 
strain hardening coefficient $\Theta \equiv x$ would be 
$P(S|\Theta) = S^{-\tau} {\cal{P}}(S/\Theta^{-d_f \nu})$, with 
${\cal{P}}\equiv\tilde{\mathcal{Z}}$,
$\tau \equiv -\lambda_z/\lambda_y$, and $\d_f \nu \equiv -\lambda_y/\lambda_x$.





\end{jpsBox}

When the external stress is raised above the yield stress, clear signs
of collective behavior arise. Dislocations stretch, rearrange, and entangle
to mediate plastic deformation, and their entanglement at least at first
raises the yield stress (work hardening, \tBox{WorkHardening}).
Furthermore, there are two indications that the plastic deformation is 
associated with an {\em emergent scale invariance} (\tBox{RG}),
with self-similar behavior on a broad range of length and time scales. 

\begin{jpsBox}{h}{Avalanches}{Avalanches}

%

\begin{minipage}{1.1\textwidth}
\centering
\includegraphics[width=\textwidth]{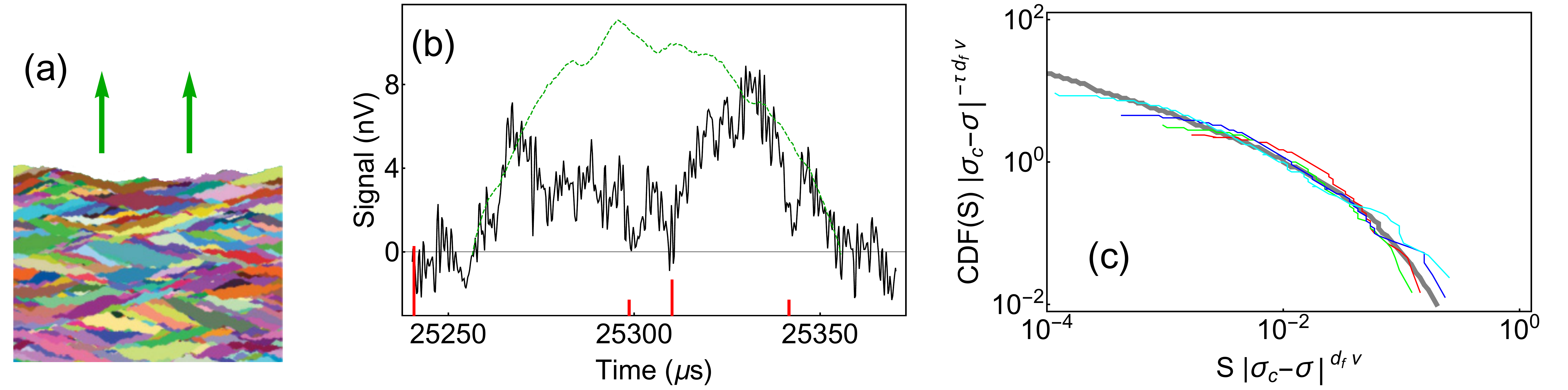}
\end{minipage}

\noindent
We call transitions from one metastable state to another under forcing
{\em avalanches} if they span a broad range of sizes. They arise not only
in plastic slip events~\cite{CsikorMWZZ04,friedman12}, but
also earthquakes~\cite{BurridgeKnopoff,bak1989earthquakes},
fracture~\cite{petri1994experimental,garcimartin1997statistical} 
and many other systems~\cite{SethnaDM01}. 
(a)~Avalanches in a front depinning transition~\cite{ChenPSZD11}
(e.g., coffee invading a napkin dipped into the cup).
(b)~A magnetic avalanche~\cite{PapanikolaouBSDZS11}, illustrating
the fractal structure in time $t$. The avalanche appears as a self-similar
sequence of smaller events, each barely triggering the next (some large trigger events marked by red bars):
here a conjunction of two medium-sized avalanches, 
which are in turn conjunctions of small avalanches, {\em etc}. The green curve
is the signal $\langle V(t) \rangle$ averaged over all avalanches
of the same duration $T$; scaling~\cite{SethnaDM01} predicts
$\langle V(t|T) \rangle \sim T^{1-d_f/z} {\mathcal{V}}(t/T)$, where $z$
is a dynamic critical exponent.
(c)~Scaling plot for dislocation avalanche sizes,
from micropillar experiments~\cite{friedman12}. The
avalanche size cutoff grew with increasing stress (different colors),
quantitatively following the mean-field prediction 
$P(S|\sigma) = S^{-\tau} {\mathcal{P}}(S/(\sigma_c-\sigma)^{-d_f \nu})$
(grey curve). Hence, avalanches can display scale-invariance in time, space, and size.

\end{jpsBox}

First, as in many rigid non-equilibrium systems, crystals respond to external
stress via dislocation {\em avalanches} (\tBox{Avalanches}).
These avalanches have been observed in bulk ice crystal
plasticity~\cite{Weiss} and micropillar plasticity of a variety of fcc and
bcc metals~\cite{Dimiduk,friedman12}. They
come in a significant range of sizes,
with a power-law probability distribution of the
net slip $P(S)\sim S^{-\tau}$.
The avalanches have a spatial fractal dimension $d_f$ less
than three~\cite{CsikorMWZZ04,WeissM03}, mostly extending along
slip bands. Such avalanches are seen
in many other rigid statistical mechanical models under forcing
(\tBox{Avalanches}).

This power-law distribution of slip sizes 
does not extend forever. Were the
distribution a pure power law, the fraction of slip produced by avalanches
larger than a size $S_0$ would be 
$\int_{S_0}^\infty S P(S|L) \d{S} \propto S^{2-\tau}|_{S_0}^\infty$; since
$1<\tau<2$ this would suggest that plastic deformation should be dominated
by the largest events. The fact that the fractal dimension $d_f$ of the
avalanches is below three provides one natural cutoff. For a sample of size
$L$, or a sample with avalanche-blocking internal structures, such as grain boundaries,
of size $L$, the biggest avalanches will be of size $L^{d_f}$. Think of
a simple model for micropillar deformation (\tBox{Precursors}),
where an avalanche induces one layer to slip over another
over an area $S$ along a slip direction. The largest avalanche will span
the pillar, so $S \sim L^2$ -- but the pillar has changed height by only a
few {\AA}ngstroms. More specifically
(see \tBox{RG}) the scaling distribution of avalanche sizes, cut off by a
length $L$ (system size, grain size, \dots), should take the form
$P(S) \sim S^{-\tau} {\cal{P}}(S/L^{d_f})$. In many cases, it is
believed~\cite{CsikorMWZZ04}
that the dominant cutoff in many cases is due not to the system
size, but to work hardening. As an
avalanche proceeds, it increases the dislocation density in its immediate
environment, raising the stress needed to produce further deformation.
%
%
%
A work-hardening cutoff is also described by a scaling form
$P(S) \sim S^{-\tau} {{\cal{P}}(S / \Theta^{-d_f\nu})}$, where 
$\Theta = \d{\sigma_Y}/\d{\epsilon}$ is the strain-hardening coefficient
(the slope of the stress-strain curve due to work hardening).

\begin{jpsBox}{h}{Cell structures \& scaling}{CellStructures}

\begin{wrapfigure}[7]{l}[0pt]{0.5\textwidth}
\label{fig:cellstructures}
\vskip -0.35truein
\null\hskip -0.05truein \includegraphics[width=0.5\textwidth]{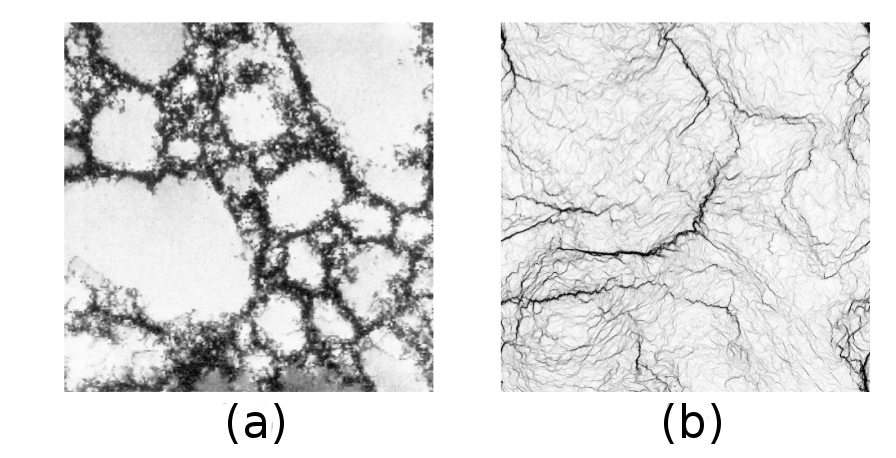}
\end{wrapfigure}


\noindent When dislocations interact they don't just form a giant ball of
spaghetti (smoothly varying distributions of highly entangled defects) but
instead form interesting cell structures, with dislocation-poor cell interiors
surrounded by dislocation-rich walls. Since these structures contain most of
the defect content of the material they dominate plastic deformation under load,
with the walls providing a rigid backbone to the softer cell
interiors~\cite{mughrabi1983dislocation}.  (a)~Just as the avalanches of
dislocations (see \tBox{Avalanches}) occur on all time-scales, these cell
structures display structures on many length-scales 
[TEM micrograph of deformed copper, from~\cite{hahner1998fractal}].
Two different methods have been used to analyze cell structures.
When analyzed using a
single length scale, universal distributions emerge that are independent of
material, loading process, or strain~\cite{hughes1998scaling}, {\em refining}
with strain instead of coarsening with time (\tBox{Nonuniversality}). 
An analysis using box-counting of the dislocation density, others find
that these cell
structures exhibit fractal morphology~\cite{hahner1998fractal}. 

(b)~Continuum models, such as that by our group~\cite{ChenCPS13}, are able to
qualitatively reproduce the complex fractal cell morphologies of real materials
using grossly simplified models of dislocation dynamics.
Our dislocation densities, misorientations,
and other physical quantities can be described using power-laws and scaling
functions. We also reproduce the universal
distributions and refinement observed by single-length scaling experimental
analyses.

\end{jpsBox}

The second indication of emergent scale invariance is the complexity of the
dislocation tangles that emerge under plastic deformation.
They are not homogeneous tangles of spaghetti -- they develop correlated
patterns with structure on many length scales. \tBox{CellStructures}
discusses a characteristic {\em cell structure} morphology found in deformed
fcc metals. It is not yet clear whether cell structures are themselves
self-similar scale invariant
fractals~\cite{hahner1998fractal,ChenCPS10,ChenCPS13} or whether
they are similar to rescaled versions of the same system at different
strains~\cite{hughes1998scaling} (like coarsening, see \tBox{Nonuniversality}),
the complex and yet patterned form of the tangles
clearly indicates the need for a theory that embraces structures on different
scales.

Systems with an emergent scale invariance -- that `look the same' at different
length scales and time scales -- are a central focus of statistical physics.
\tBox{RG} briefly summarizes key ideas distilled from
an enormous, sophisticated literature on a wide variety of systems. At a 
continuous transition between two qualitative behaviors ({\em e.g.} stuck
to sliding as stress is tuned), many systems will exhibit 
scale-invariant fluctuations with regions small and large alternating
between the two qualitative behaviors. 
{\em Renormalization group} methods use this emergent self-similarity
to predict power-law behavior of functions with one control parameter
(like $P(S) \sim S^{-\tau}$ above) and predict scaling forms for properties
depending on more than one parameter 
(like $P(S|L) \sim S^{-\tau} {\cal P}(S/L^{d_f})$).

\begin{jpsBox}{h}{Fracture in disordered media}{Fracture}

\begin{wrapfigure}{L}{.7\textwidth}
	\vspace{-1.4em}
	\includegraphics[width=.7\textwidth]{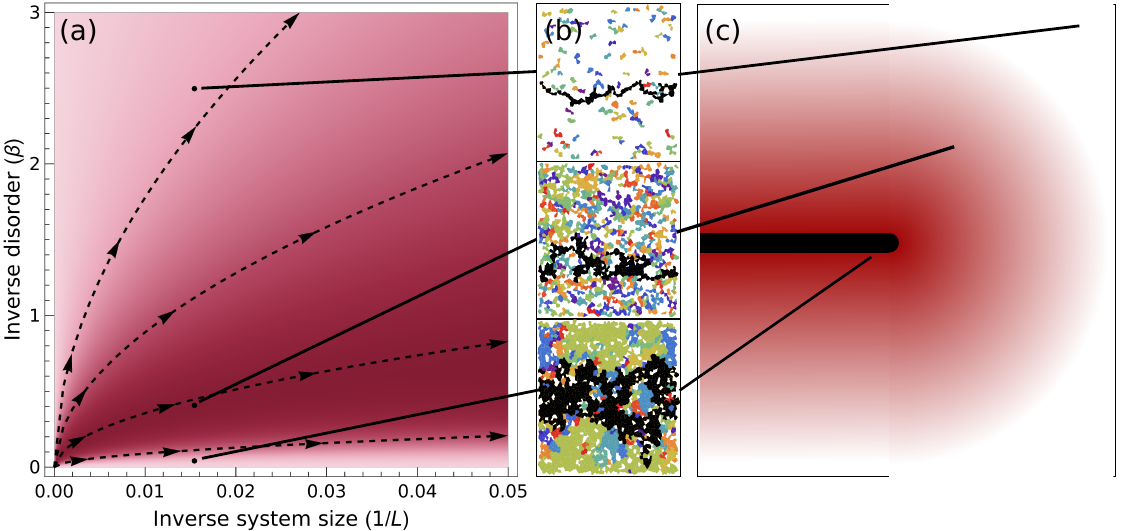}
	\vspace{-2.7em}
\end{wrapfigure}
\noindent
Quasi-brittle materials like concrete are disordered, with a
broad distribution of microscopic strengths~\cite{BertalanSSZ14}
parameterized by inverse disorder $\beta$. Low disorder or large
sample size correspond with brittle, nucleated fracture 
[(a)~upper left, (b) top],
while high disorder or small size correspond with uncorrelated
percolation-like fracture [(a)~lower boundary, (b)~bottom].
The crossover between these regimes shows large avalanches [(b)~middle],
whose mean second moment is indicated by red intensity.
The RG flows [(a)~arrows] predict that the distribution of avalanche sizes
$P(s|\beta,L)$ has a crossover scaling form
$P(s|\beta,L)=s^{-\tau}{\mathcal{P}}(\beta L^{1/\nu_f},sL^{-1/\sigma\nu_f})$;
no phase boundary separates the two regimes.
(c)~This scenario may also apply to the damage (schematically indicated by red
intensity) in the
region around a growing
crack for an infinite system. Near the growing crack tip, stress is
high and the material is reduced to rubble (percolation), while far
from the crack tip the stress is small and the material undamaged. The
same RG crossover scaling analysis should allow us to develop
a quantitative scaling theory of this damaged
{\em process zone}~\cite{JaronSSnn}.

\end{jpsBox}

Our first example is drawn from another materials
challenge, using scaling ideas to explain various aspects of fracture. 
First, much attention has been
spent on fractal analyses of the resulting fracture
surfaces~\cite{alava2006morphology,zapperi2005crack,%
alava2006statistical,bouchaud1990fractal,bouchaud1997scaling,ponson2006two,%
morel2000scaling,maaloy1992experimental,hansen2003origin}.
Two rival theories of crack growth
in disordered materials~
materials~\cite{schmittbuhl2003roughness,laurson2010avalanches,schmittbuhl1995interfacial,rosso2002roughness}
each turned out to describe the fractal height fluctuations
observed in
experiments~\cite{bouchaud1990fractal,bouchaud1997scaling,maaloy1992experimental,mecholsky1989quantitative,mecholsky1988self,mecholsky1991relationship,tsai1991fractal,schmittbuhl1997direct,schmittbuhl1995scaling,schmittbuhl1993field},
one at short distances and one at longer distances. 
A crossover scaling
theory~\cite{santucci2007statistics,ChenZS15} produces a unified
description of experimental
height correlations on
intermediate length scales. Second, in brittle materials, crack nucleation
is studied with {\em extreme value
statistics}~\cite{talreja2013probability,jayatilaka1977statistical,phoenix1973asymptotic} -- the distribution
of failure strengths is described by universal Gumbel and Weibull distributions,
which recently have been
viewed as renormalization-group fixed
points~\cite{gyorgyi2010renormalization,gyorgyi2008finite}.

\begin{marginnote}[]
\entry{Process zone}{The damaged zone near the tip of a crack, where
linear elastic theory breaks down. For concrete, this can be nearly 
a meter in size.}
\end{marginnote}

The third scaling approach to fracture describes 
{\em quasi-brittle} materials like
concrete, in which the strength of local regions is highly disordered
(\tBox{Fracture}).
Similar to dislocation avalanches in plastically deformed crystals,
one observes a power-law distribution of microcrack 
avalanches preceding the eventual
rupture~\cite{salminen2002acoustic,koivisto2007creep,hemmer1992distribution,petri1994experimental,garcimartin1997statistical}. These are due to {\em finite-size
criticality}~\cite{ShekhawatZS13}; large avalanches in a crossover
region between two regimes neither of which have bulk avalanches.
At small sizes and large disorder, the bonds break in a percolation-like
fashion (one at a time); at large sizes a single dominant crack is nucleated in 
a rare event. Our finite-size scaling analysis should be generalizable
into a crossover scaling theory for the process zone near the crack tip
in quasi-brittle materials (\tBox{Fracture}c). Combining these ideas
with a scaling theory of plasticity (as described here) could allow for
a crossover theory for the process zone for ductile fracture as well.

There are two classes of models that provide a particularly clear connection
to the physics of plastically deformed crystals: the theory of {\em depinning}
(described in \twoBoxes{Depinning}{MesoscalePlasticity}),
and the theory of {\em jamming} (in \tBox{Jamming}). Both describe
the response of a rigid system to external shear, via avalanche-like
rearrangements.

\begin{jpsBox}{h}{Depinning and criticality}{Depinning}

\begin{minipage}[t]{0.4\textwidth}
\centering
\begin{tabular}{| l | c | c | c |}
\hline System & $d$ & $d^\prime$  & $N$ \\ \hline\hline
Planar cracks	& 1 & 2 & 1 \\ \hline
CDWs 		& 3 & 3 & 1 \\ \hline
Vortices 	& 1 & 3 & 2 \\ \hline
Droplets 	& 1 & 2 & 1 \\ \hline
Magnetic 	& 2 & 3 & 1 \\ \hline
Plasticity 	& 3 & 3 & 12 \\ \hline
\end{tabular}
\end{minipage}
\hspace{0.2cm}
\begin{minipage}[t]{0.7\textwidth}
\centering
\raisebox{-.5\height}{\includegraphics[width=\textwidth]{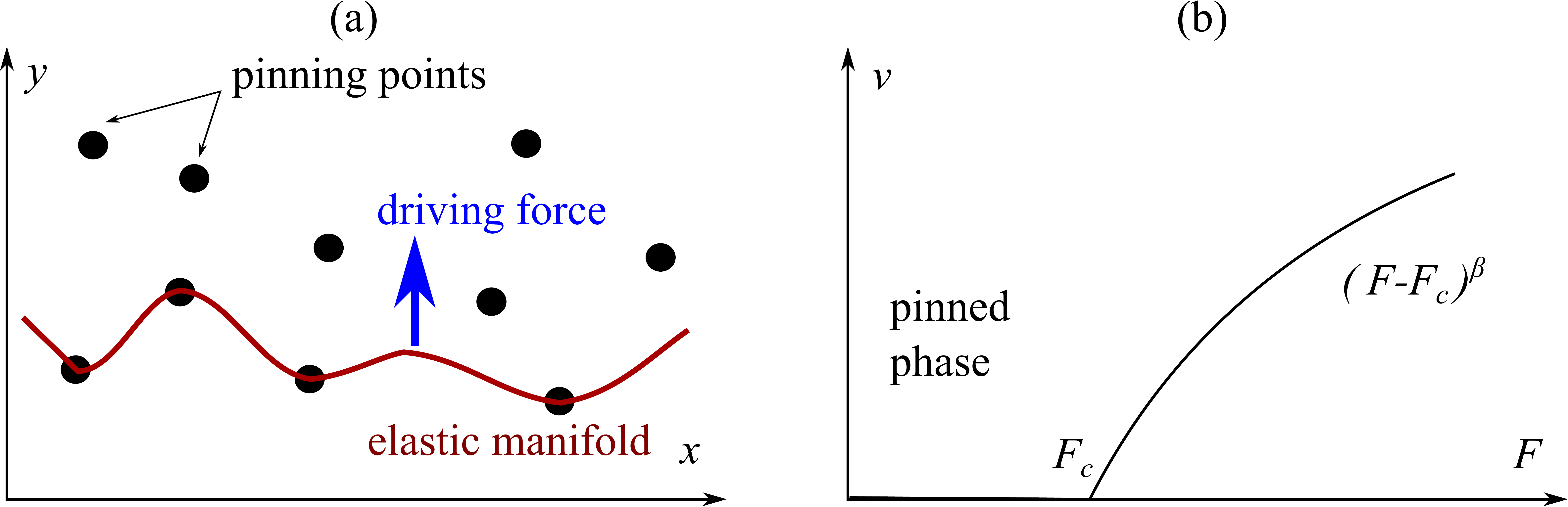}} %
\end{minipage}

\vspace{0.2cm}

\noindent
(Table) Depinning transitions describe the jerky movement of surfaces pinned by disorder
as they is dragged; they have been used to describe everything from planar cracks
to superconducting vortices to raindrops on windshields~\cite{fisher98}.
(a) A $d$-dimensional manifold is stuck on
dirt in a $d^\prime$-dimensional space, and can move along $N$ different
directions. As the force increases, the
system exhibits avalanche-like rearrangements of local regions, shifting
between metastable states stuck in the dirt, until at a critical force
the system starts sliding. (b) Universal power laws and scaling behavior
in the velocity and velocity autocorrelation above depinning, and in
the avalanche sizes, durations, shapes, and spatiotemporal correlations
are explained using the emergent scale invariance at the depinning
threshold. Fcc crystal plasticity is an example of a $d=d'=3$ depinning transition,
with $N=12$ slip systems yielding under an external load
(although $d=d^\prime=2$ and $N=1$ in the simplest
model~\cite{zaiser05, zaiser06}). Depinning transitions often have
self-limiting terms that allow them to self-organize near critical
points; work hardening plays this role for plasticity.

\end{jpsBox}

\begin{jpsBox}{h}{Mesoscale plasticity}{MesoscalePlasticity}

\begin{wrapfigure}[4]{L}{0.4\textwidth}
\vskip -0.9cm
\begin{align}
& \frac{1}{B} \partial_t \gamma (\bm{r}) = \sigma_{\text{ext}} + \sigma_{\text{int}} (\bm{r}) + \delta \sigma (\bm{r}, \gamma) \nonumber\\
& \sigma_{\text{int}} (\bm{k}) = -\frac{G}{\pi (1-\nu) \gamma (\bm{k})} \frac{k_x^2 k_y^2}{|\bm{k}|^4} 
\end{align}
\end{wrapfigure}

\noindent
This mesoscale plasticity evolution equation [from~\cite{zaiser05}] is 
typical of depinning: 
the evolution of the manifold $\gamma$ depends on an external 
driving force $\sigma_{\text{ext}}$, a nonlocal force $\sigma_{\text{int}}$ due 
to the rest of the manifold, and a force $\delta\sigma(\bm{r},\gamma)$ due 
to dirt and other dislocations at $\bm{r}$. Here $\gamma$ is the shear
strain (up to twelve components), $G$ the shear modulus,
$\nu$ the Poisson ratio, and $B$ a viscoplastic rate coefficient. The nonlocal
kernel given in Fourier space by the lower equation is characteristic of
all elastic materials -- both crystalline and amorphous~\cite{zaiser06, talamali11}. Plasticity is
different from most depinning problems for two reasons. First, the kernel is
not convex (positive and negative in different directions), making most
of the analytic methods challenging. Second, dislocations tangle among
themselves even without dirt. Like glasses and jammed systems, they 
generate their own disorder as they evolve.

\end{jpsBox}

Depinning transitions describe motion in the presence of dirt, as for dislocations pinned primarily on second phase precipitates.  First, in fields like magnetic hysteresis and noise, depinning theories provide a comprehensive and compelling framework for understanding the experimental behavior~\cite{ZapperiDurinReview} and are pursued using sophisticated functional renormalization-group methods~\cite{FunctionalRGMagnetism1,FunctionalRGMagnetism2,FunctionalRGMagnetism3,FunctionalRGMagnetism4,FunctionalRGMagnetism5}.  Second, in fields like earthquakes things are more controversial; the data does not discriminate so well between different models, but there is a broad consensus that scaling approaches are important and useful~\cite{DahmenEarthquakes3,DahmenEarthquakes1,BaroCIPSSSV13}; Within the scaling theories, depinning models have had notable success~\cite{FisherDRB97,DahmenEarthquakes1,chen91}, but remain in competition with other types of models~\cite{BurridgeKnopoff,DahmenEarthquakes4,MyersCarlsonEarthquakes}. Part of the challenge is that the long-range interactions through the earth’s crust make the theory mean-field~\cite{ErtasD}, allowing many different microscopic theories to yield the same behavior. Finally, many of the statistical theories of plasticity in  crystals are depinning theories (see \tBox{MesoscalePlasticity}).

\begin{jpsBox}{h}{Jamming}{Jamming}

\begin{wrapfigure}[10]{L}{0.72\textwidth}
\vspace{-0.5cm}
\begin{minipage}{.3\textwidth}
 \hspace{-3cm}
 \includegraphics[height=4cm]{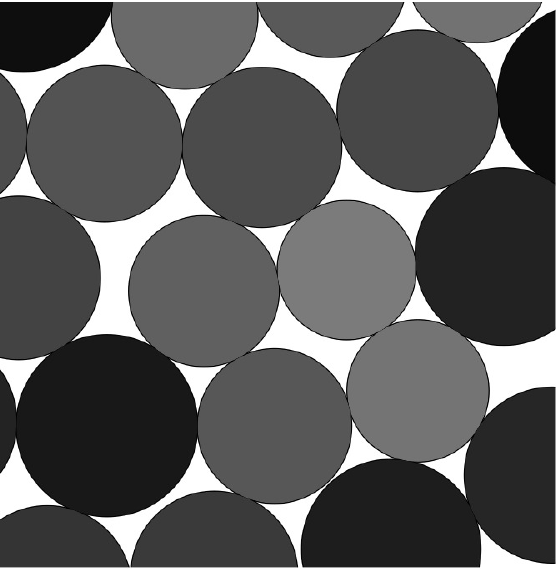}
\end{minipage}%
\begin{minipage}{.3\textwidth}
 \hspace{-2.2cm}
  \includegraphics[height=4cm]{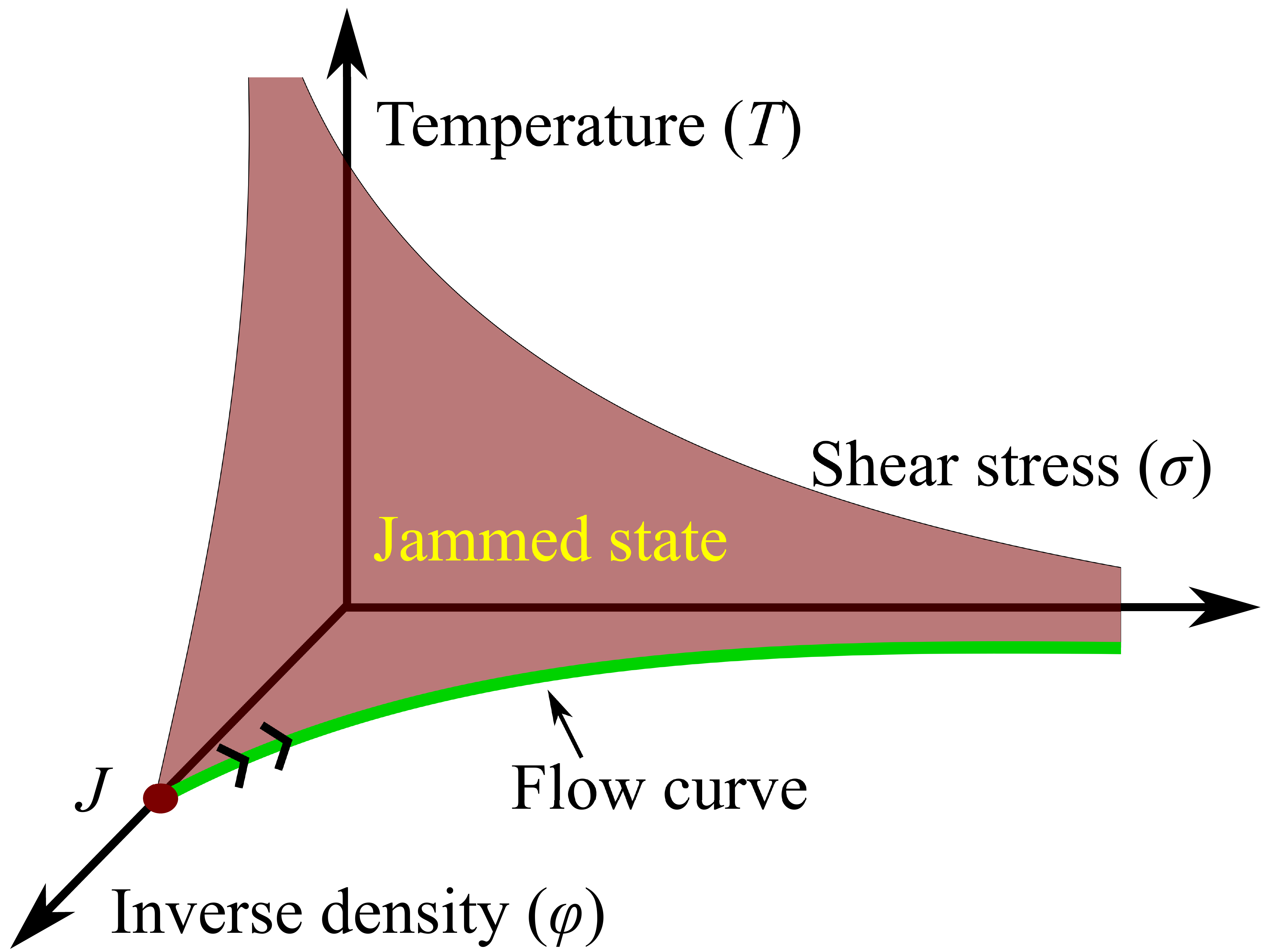}
\end{minipage}
\end{wrapfigure}

\noindent
Jamming~\cite{liu10,liu11} attempts to describe the onset of rigidity in
molecular glasses, granular media, colloids, pastes,
emulsions, foams~\cite{liu11}, dislocation systems~\cite{ispanovity14},
and biological tissues~\cite{bi16}. At low temperature, low stress,
and high density the system is stuck in a jammed phase---a disordered 
solid state that resists shear. The \emph{jamming
point} $J$ controls the critical behavior of the system. In contrast
to depinning, where dirt is explicitly part of the model,
the disorder in a jammed system is frozen in as it jams -- just as dislocations
in crystals tangle themselves up as they 
evolve~\cite{tsekenis11,miguel02,ispanovity14}.
At point $J$ one finds the
power laws, scaling collapses, and diverging length scales~\cite{liu10,liu11}
that are characteristic of systems with emergent scale invariance 
(\tBox{RG}), although no coarse-graining approach has yet
been developed. A recent scaling {\it{ansatz}}~\cite{GoodrichLS16} argues
that the free energy has the scaling form
$F(\Delta \varphi, P, \sigma, T) = \Delta\varphi^{2}
\mathcal{F}_0 ( P / \Delta \varphi, \sigma / \Delta \varphi^{5/4}, T /
\Delta \varphi^{2})$, where (as usual) the arguments of the scaling
function are invariant under the presumed renormalization-group flow
(\tBox{RG}), with mean-field rational critical exponents 
in three dimensions. In particular, a point near $J$ along the zero-temperature
yielding
boundary (green curve in the $\varphi$-$\sigma$ plane), when coarse-grained,
must also be a system at its yield point, farther from point $J$. This
implies that the invariant combination $\sigma / \Delta \varphi^{5/4}$
is invariant along the yielding curve, so $\sigma_\mathrm{flow} \propto
\Delta \varphi^{5/4}$~\cite{hatano08,tighe10,dinkgreve15}. Jammed states may
develop equilibrium-like properties; they seem to maximize
entropy (become equally probable~\cite{martiniani16}), and to have `thermodynamic'
descriptions (see Refs.~\cite{MakseKurchan,AbateD08,BerthierB02,OnoODLLN02}
and \tBox{CompactivityAngoricity}). Perhaps the weird memory effects of
{\bf Section~\ref{sec:Memory}} are erased by the large fluctuations near critical
points.

\end{jpsBox}

Jamming transitions~\cite{liu10,liu11}
describe the elastic response of systems near the point
where the constituent particles first assemble into a rigid network.
The hardness of clean single crystals is primarily due to dislocation 
entanglement with other dislocations -- the disordered pinning is not
static, but dynamically evolves as the dislocations `jam' together, making
the analogy with the jamming critical point perhaps more
plausible~\cite{miguel02,tsekenis11,ispanovity14}. 

In plastically deformed crystals, what is the critical point? What
corresponds to the depinning force in \tBox{Depinning}
or point $J$ in \tBox{Jamming}? Let us consider a system where the slope
of the work hardening curve $\Theta = \d\sigma_Y / \d\epsilon$
controls
the cutoff in the avalanche size distribution, and presumably also the 
cutoff for the fractal self-similarity in the spatial morphology. The
critical point $\Theta=0$ (which flows to self-similar fixed point under the
renormalization group in \tBox{RG}) corresponds to a material which does not
work harden with increasing strain. 

Such a material would either continue to deform at constant stress (perhaps
related to {\em superplasticity}~\cite{SuperPlastic}), or the
material begins to weaken with further stress (as happens in {\em metallic
glasses}, leading to shear band failure~\cite{budrikis13,sandfield15}).
Evidence for a `plain old' critical point at a stress $\critstress$
has been seen in 
micropillar experiments~\cite{friedman12} (\tBox{Avalanches}(c)),
where the avalanche sizes diverge at a $\critstress$
that we call the critical failure stress. They find 
quantitative agreement with a mean-field scaling theory, for 
a traditional critical point, with a scaling form for the avalanche sizes
$P(S|\stress) \sim S^{-\tau} {\cal P}(S/(\critstress-\stress)^{-d_f \nu})$.

\begin{marginnote}[]
\entry{Recrystallization}{An abrupt change in the morphology of a 
work-hardened crystal, where the dislocations become so dense that they
spontaneously annihilate, mediated by a large
angle grain boundary that sweeps through the system.}
\end{marginnote}

\begin{jpsBox}{h}{Mean field theory \& plasticity}{MeanFieldPlasticity}

\begin{wrapfigure}{L}{0.21\textwidth}
\vskip -0.5cm
\begin{minipage}{.21\textwidth}
\begin{picture}(4,56)
\put(0,0){\includegraphics[width=.7\linewidth]{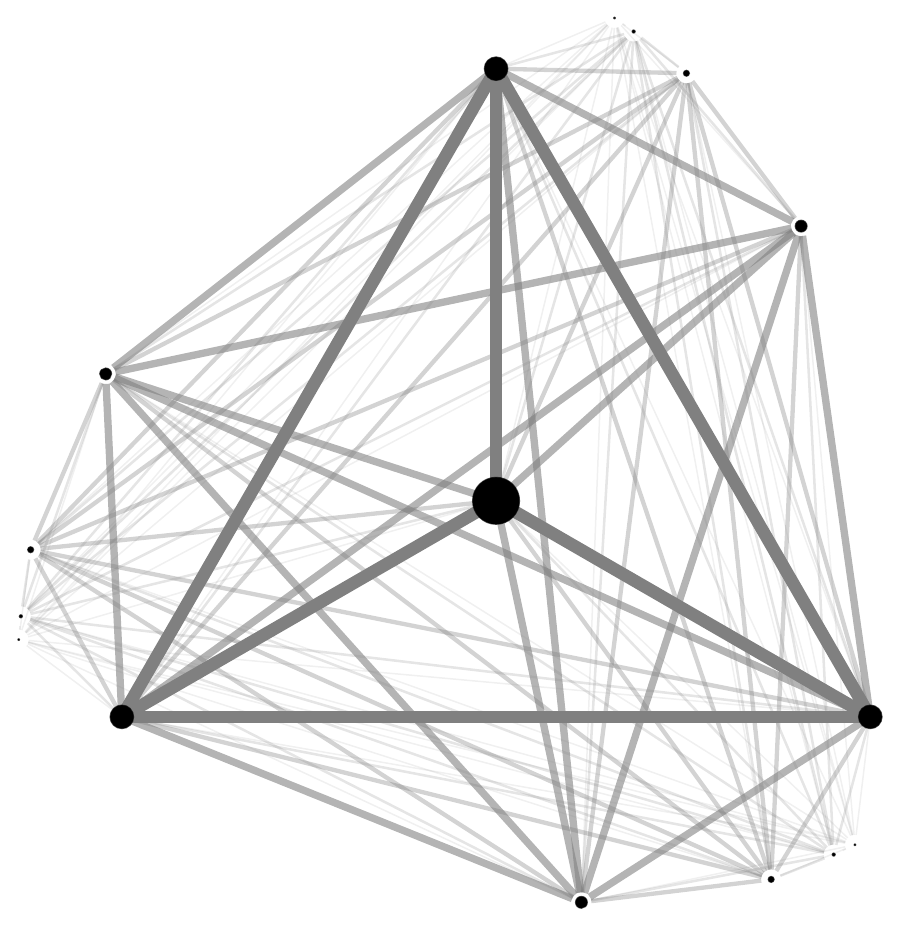}}
\put(0,50){(a)}
\end{picture}

\begin{picture}(4,55)
\put(0,43){(b)}
\put(0,-10){\hspace{.7cm}\includegraphics[angle=30,width=.5\linewidth]{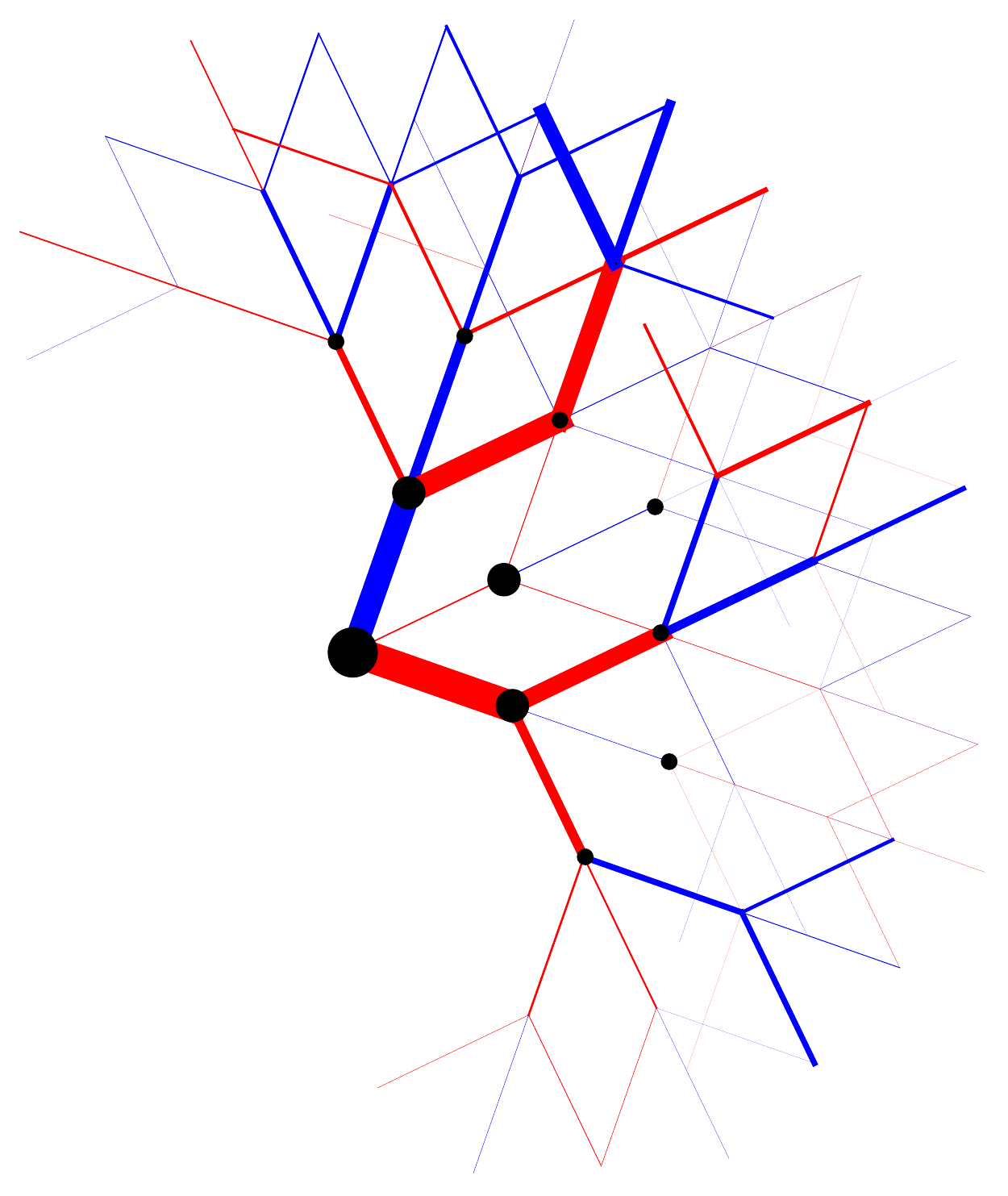}}
\end{picture}

\begin{picture}(4,60)
\put(3,-5){\includegraphics[width=\linewidth]{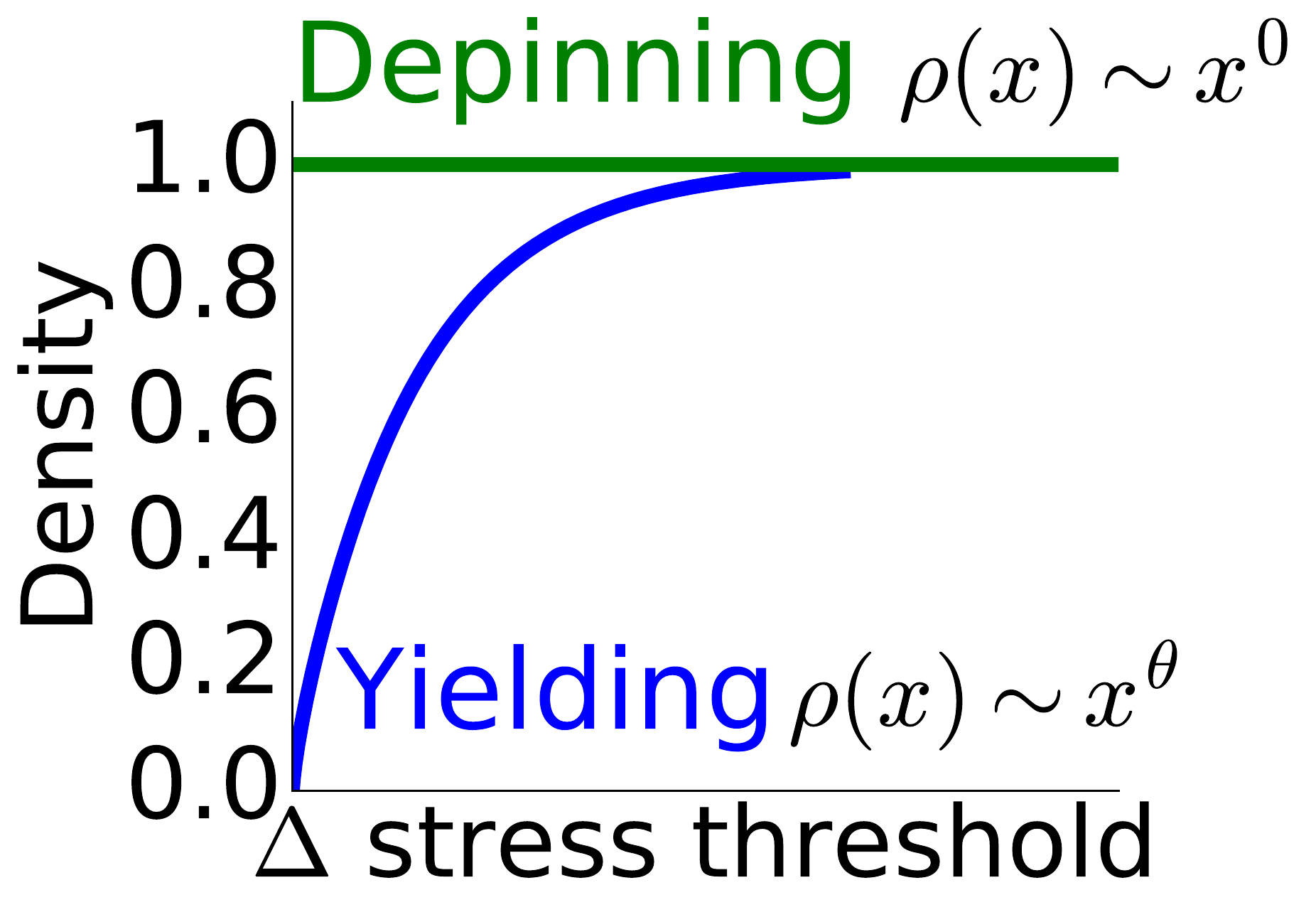}}
\put(0,45){(c)}
\end{picture}
\end{minipage}
\end{wrapfigure}

\noindent
Mean field theories describe systems in high dimensions or with
long-range forces. Some assume each degree of freedom feels the mean
of other sites like in a fully connected hypertetrahedron~(a); others
sit on an infinite branching tree (Bethe lattice, b). With a plastic
rearrangement, the anisotropic stress increases the loading on some
neighbors and decreases it on others. Some mean-field plasticity models
focus on the decrease but ignore the anisotropy
\cite{Pazmandi:2000ep,WyartMeanField}, others argue that the anisotropy
allows one to ignore the decreases \cite{zaiser06,Dahmen:2009iz}.
(c)~For systems of the former type, the density of sites close to their
failure threshold must be suppressed due to past fluctuations
crossing the threshold, forming a {\em pseudogap} scaling as 
$(\sigma_{\mathrm{c}}-\sigma)^\theta$. This yields a value of
$\tau \le 3/2$ that depends on the long-range tail in the coupling
distribution. For the
latter, no pseudogap appears, and the distribution of avalanche sizes
scales as $P(S) \sim S^{-\tau}$ where $\tau=3/2$.  A wide variety of
experiments are consistent with
$\tau=3/2$~\cite{PapanikolaouBSDZS11,Sun:2010bh,Weiss,Antonaglia:2014gc}. 
In contrast, some simulations suggest that
$\tau\approx 1.3$~\cite{Liu:2016ft,Budrikis:2015tu,Salerno:2013dz}.

\end{jpsBox}


Much of the statistical physics literature describes crystal plasticity
instead as a {\em self-organized} critical point~\cite{SOCReview}.
Depinning systems like earthquakes are thought to exhibit emergent scale 
invariance without fine-tuning to a critical point because the natural
microscopic velocities (tectonic plates drift centimeters/year) are tiny
compared to the mean velocities during avalanches~\cite{SethnaDM01}.
In many macroscopic materials, work hardening under increasing stress 
or strain does not peak at a critical stress or strain.%
   \footnote{Instead, it terminates in a qualitative way -- fracture
   (breaking in two), amorphization (where the crystalline correlations
   drop to the atomic scale), or perhaps 
   {\em recrystallization}~\protect\cite{rollett2004recrystallization}.}
Also, the natural scale of the work hardening coefficient $\Theta$ 
is comparable to the elastic modulus; one prediction~\cite{CsikorMWZZ04}
for the cutoff of the avalanche size distribution is a slip of%
   \footnote{We define $S$ to be the net slip. Csikor defines their avalanche
   size $s$ to be the strain jump, which is $b S/L^3$, since a slip of size 
   $L^2$ causes the material to shrink by around a Burgers vector 
   $\delta L/L \sim b/L$. Csikor's cutoff at constant load $s_0 = bE/L\Theta$
   thus becomes a cutoff in our notation of $S_0 = L^2 E/\Theta$.}
$S_0 = L^2 E/\Theta$
-- small compared to macroscopic scales, but large compared to atomic scales
for large system size $L$.
Thus the natural cutoff scale due to work hardening for macroscopic samples
is tiny, just as the velocities of earthquake fronts are tiny, leading to a
self-organized critical point.%
   \footnote{Here there are two different relevant variables at the fixed 
   point, $1/L$ and $\Theta$. The predicted scaling form for the avalanche size
   distribution 
   $P(S|\Theta,L) \sim S^{-\tau} {\mathcal{P}}(S/L^{d_f},L/\Theta^{-\nu})$
   gives the distribution used by Csikor {\em et al.} 
   $P(S) =  S^{-\tau} \e^{(S/S_0)^2}$ if we assume 
   ${\mathcal{P}}(X,Y) = \exp\left((X Y^{1/\nu})^2\right)$ and
   $d_f-1/\nu = 2$. Note that this makes sense only if $d_f > 2$,
   different from that proposed by Csikor {\em et al.}, but compatible
   with $d_f\sim2.5$ found by Weiss~\cite{WeissM03}.}

A key test for scaling theories of plasticity is the exponent $\tau$ giving
the power law for the avalanche size distribution. Many experiments on
crystalline and amorphous plasticity can be
fit using a value $\tau\sim 3/2$~\cite{Dahmen:2009iz}. While older simulations seemed to agree with this value~\cite{csikor2007dislocation,zaiser2005yielding}, newer simulations
observe smaller values of $\tau \sim
1.25-1.4$~\cite{budrikis13,Liu:2016ft,Salerno:2013dz,Budrikis:2015tu,Simulations}.%
    \footnote{One set
    of experiments~\cite{PapanikolaouDCSUWZ12} showed a rate-dependence yielding
    values of $\tau > 1.5$ for slow deformations; these experiments have been
    modeled as a quasi-periodic approach to a critical point punctuated by
    quasi-periodic system-spanning events~\cite{jagla14}.} 

It is not clear at this time what
governs these variations, but one intriguing hint is provided by two different
{\em mean-field theories} for plasticity (\tBox{MeanFieldPlasticity}).
A local rearrangement of atoms at a site will yield a net plastic slip,
producing a long-range, anisotropic change in the stress. This {\em elastic
dipole} adds to the imposed stress in some directions, and relieving the
stress in others, often triggering disconnected pieces of the avalanche
elsewhere in the system. Some mean-field theories~\cite{Dahmen:2009iz,zaiser06} 
have ignored the variation in sign of this interaction, arguing
that the directional anisotropy makes the positive interactions
within a slip band or glide plane the only important ones~\cite{FisherDRB97}.
This
leads to a theory where every site feels a monotonically increasing external
stress, which in mean-field gives the exponent $\tau = 3/2$.
Others~\cite{ZimanyiSpinGlass,WyartMeanField} have ignored the directional
anisotropy, and argue that the 
random positive and negative changes in stress lead to a kind of 
anomalous diffusion in the distance of each site from its local yielding
stress. Since any site that crosses its yielding stress will trigger an
avalanche and disappear from the active list, this carves a hole
out of the distribution of sites near their critical stress. This 
leads to fewer, larger avalanches and a lower value for $\tau$. 

We end with a cautionary note, particularly addressed to the statistical
physics community. Emergent scale invariance in physics leads not only
to power laws and scaling functions, it also leads to {\em universality} --
any two critical systems which flow to the same fixed point in
\tBox{RG} will share the same long-wavelength behavior. Thus
the Ising model quantitatively describes both the disappearance of 
magnetism with increasing temperature in some magnets, and the 
disappearance of the density difference between liquid and vapor along
the coexistence line as temperature and pressure are increased. 
In contrast, success in materials physics has historically rested upon 
attention to
materials-specific details. Many metallurgists focus their careers
on specific materials. Aluminum alloys, steels, and titanium superalloys
are each worlds unto themselves. Are the scaling methods of statistical
physics doomed in the attempt to describe the bewildering variety of
anisotropic slip systems, dislocation mobilities, cross slip, pinning, \dots?

\begin{jpsBox}{h}{Non-universality}{Nonuniversality}


\begin{wrapfigure}[7]{l}[0pt]{0.4\textwidth}
\vskip -0.15truein
\null\hskip -0.05truein \includegraphics[width=0.4\textwidth]{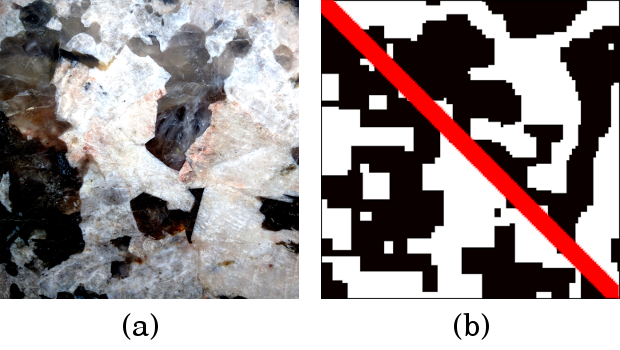}
\end{wrapfigure}

\noindent
(a)~Polycrystalline granite. When magma slowly cools below the Earth's
surface, it {\em coarsens} into different component crystals, forming
granite. Scaling theories that describe coarsening have inverse time as
a relevant variable.
When $r \to r/b$ times scales as $t \to
t/b^3$ (\tBox{RG})~\cite[sec.~11.4.1]{Sethna06}.
Thus salad dressing and the Ising model near $T_c$
are both described by the same correlation function $C({\mathbf{r}},t) =
\mathcal{C}(r/t^{1/3})$ since coarsening in isotropic systems is
\textit{universal}. In 1999, Rutenberg and Vollmayr-Lee noted that this
is {\em not true} of anisotropic crystalline systems, or even the Ising
model away from $T_c$; although a scaling form exists, it is dependent
on the anisotropic surface tensions and interface
mobilities.~\cite{RutenbergAnisotropic} Panel~(b) shows a comparison of
coarsening in the 3D Ising model (upper triangle) to the same system
with additional weak antiferromagnetic next-neighbor bonds (lower
triangle). The added interaction slows coarsening to $r\sim\log(t)$ and
introduces a preference for flat boundaries aligned with the lattice
directions~\cite{ShoreHS92}, producing facets like those in granite~(a).
Although physicists are fond of universality, 
the world is filled with a bewildering variety of rocks and dislocation
tangles. Many materials-dependent properties likely must be treated as
``relevant variables" in our eventual scaling theory of plasticity, as
they are for coarsening.

\end{jpsBox}

Here we remind our statistical colleagues of a simple but profound observation
by Rutenberg and Vollmayr-Lee some decades ago (\tBox{Nonuniversality}),
that our scaling theory of {\em coarsening} is an example where the 
`universality class' is parameterized by {\em entire anisotropic functions}
of angle. 
There is no good reason why this kind of strong material dependence 
should be incompatible with scaling and criticality. Also, there likely 
will be several `universality classes' depending upon the active mechanisms
for plastic deformation; systems with one active slip system behave
very differently from the same material at later stages where multiple
systems and cross slip become active. Our attempts to form a scaling
theory must not only focus on what makes plastic deformation 
the same among materials composition and processing history, but 
also what makes each system different.

\section{Conclusions}
\label{sec:Conclusion}

\begin{summary}[SUMMARY POINTS]
\begin{enumerate}
\item Crystals are a challenge for nonequilibrium statistical mechanics.
Our methods addressing disorder, long-range forces, constrained dynamics,
and anisotropic interactions must be combined to address dislocation
entanglement in crystal plasticity.
\item There are many rigid statistical mechanical systems with behavior
closely analogous to the yield stress and work hardening seen in crystal
plasticity. Upon unloading, these simpler systems are left in configurations
that are rare among the possible metastable states, encoding the material
history in interesting ways.
\item The thermal and mechanical history of a crystal thus is probably encoded
in its morphology. This need not prevent a phenomenological theory from
effectively encapsulating its future macroscale behavior.
\item Avalanche dynamics and the morphologies of dislocation tangles provide
clear evidence that plastically deformed crystals exhibit an emergent
scale invariance.
\item Strong analogies can be made between plasticity avalanches and 
scaling behavior seen in fracture, depinning, jamming, and 
even dilute colloidal suspensions under shear. 
\item Crystal plasticity is immensely complex and strongly material and
history dependent. A successful renormalization-group scaling
theory of plasticity in crystals will depend
on far more details about the microscopic behavior than has been typical in
previous systems exhibiting scale invariance.
\end{enumerate}
\end{summary}


\section*{DISCLOSURE STATEMENT}
The authors are not aware of any affiliations, memberships, funding, or financial holdings that
might be perceived as affecting the objectivity of this review. 

\section*{ACKNOWLEDGMENTS}
We thank Bulbul Chakraborty, Karen Daniels, Andrea J. Liu, and M. Lisa Manning for extensive consultation. We also thank Paul Dawson, Ryan Elliott, Susan Coppersmith, James Jenkins, and Ellad Tadmor for kindly providing references and/or permission to reprint figures. JPS, MKB DBL and AR were supported by the Department of Energy through Grant No. DE-FG02-07ER46393. LXH, JPK, EDL and KNQ were supported by the National Science Foundation through Grant No. NSF DMR-1312160. KD is thankful for the support from NSF through Grant No. NSF CBET 1336634, and also thanks the Kavli Institute of theoretical Physics for hospitality and support through Grant No. NSF PHY11-25915. CPG is supported by the National Science Foundation through the Harvard Materials Research Science and Engineering Center DMR1420570, and the Division of Mathematical Sciences DMS-1411694. JRG and XN are grateful to the U.S. Department of Energy (DOE) through J.R.G.Õs Early Career Research Program under grant DE-SC0006599. LXH was supported by a fellowship from Cornell University. EDL thanks support by the National Science Foundation through the fellowship No. NSF GRFP (DGE-1650441). KNQ was supported by the Natural Sciences and Engineering Research Council of Canada. DZR thanks support by the Bethe/KIC Fellowship, and the National Science Foundation Grant No. NSF DMR-1308089. AS acknowledges support from the Miller Fellowship by the Miller Institute for Basic Research in Science, at UC Berkeley. SZ is supported by the European Research Council advanced grant SIZEFFECTS.


\begin{thebibliography}{167}
\expandafter\ifx\csname natexlab\endcsname\relax\def\natexlab#1{#1}\fi

\bibitem{Chandler}
Chandler D. 1987.
Introduction to modern statistical mechanics.
Oxford University Press

\bibitem{Forster}
Forster D. 1995.
Hydrodynamic fluctuations, broken symmetry, and correlation functions.
Advanced book classics. Perseus Books

\bibitem{Martin}
Martin PC. 1968.
Probl\`eme \'a n corps (many body physics).
Gordon and Breach, New York,  37---136

\bibitem{Landau}
Landau L, Lifshitz E. 2013.
Statistical physics.
No. v. 5. Elsevier Science

\bibitem{TadmorElliott}
Tadmor EB, Miller RE, Elliott RS. 2012.
Continuum mechanics and thermodynamics: from fundamental concepts to governing
  equations.
Cambridge University Press

\bibitem{PWAIllCondensed}
Anderson PW. 1984.
Lectures on amorphous systems

\bibitem{Sethna06}
Sethna JP. 2006.
Statistical mechanics: Entropy, order parameters, and complexity,
  \url{http://www.physics.cornell.edu/sethna/StatMech/}.
Oxford: Oxford University Press

\bibitem{RougheningDynamics}
Chui ST, Weeks JD. 1978.
Dynamics of the roughening transition.
\textit{Phys. Rev. Lett.} 40:733--736

\bibitem{LiarteBMKS16}
Liarte DB, Bierbaum M, Mosna RA, Kamien RD, Sethna JP. 2016.
Weirdest martensite: {S}mectic liquid crystal microstructure and
  {W}eyl-{P}oincar\'e invariance.
\textit{Phys. Rev. Lett.} 116:147802

\bibitem{VeggeSCJMR01}
Vegge T, Sethna JP, Cheong SA, Jacobsen KW, Myers CR, Ralph DC. 2001.
Calculation of quantum tunneling for a spatially extended defect: The
  dislocation kink in copper has a low effective mass.
\textit{Physical Review Letters} 86:1546--1549

\bibitem{hull11}
Hull D, Bacon DJ. 2011.
Introduction to dislocations, vol.~37.
Elsevier

\bibitem{NiExpts}
Ni X, Greer JR. 2016.
Unpublished

\bibitem{BulatovMultiJunction}
Bulatov VV, Hsiung LL, Tang M, Arsenlis A, Bartelt MC, et~al. 2006.
Dislocation multi-junctions and strain hardening.
\textit{Nature} 440:1174--1178

\bibitem{Taylor34}
Taylor GI. 1934.
The mechanism of plastic deformation of crystals. part i. theoretical.
\textit{Proceedings of the Royal Society of London A: Mathematical, Physical
  and Engineering Sciences} 145:362--387

\bibitem{budrikis13}
Budrikis Z, Zapperi S. 2013.
Avalanche localization and crossover scaling in amorphous plasticity.
\textit{Phys. Rev. E} 88:062403

\bibitem{sandfield15}
Sandfeld S, Budrikis Z, Zapperi S, Castellanos DF. 2015.
Avalanches, loading and finite size effects in 2d amorphous plasticity: results
  from a finite element model.
\textit{Journal of Statistical Mechanics: Theory and Experiment} 2015:P02011

\bibitem{dieter1986mechanical}
Dieter GE, Bacon DJ. 1986.
Mechanical metallurgy, vol.~3.
McGraw-Hill New York

\bibitem{bi15}
Bi D, Henkes S, Daniels KE, Chakraborty B. 2015.
The statistical physics of athermal materials.
\textit{Annual Review of Condensed Matter Physics} 6:63--83

\bibitem{edwards89}
Edwards S, Oakeshott R. 1989.
Theory of powders.
\textit{Physica A: Statistical Mechanics and its Applications} 157:1080 -- 1090

\bibitem{martiniani16}
{Martiniani} S, {Schrenk} KJ, {Ramola} K, {Chakraborty} B, {Frenkel} D. 2016.
{Are some packings more equal than others? A direct test of the Edwards
  conjecture}.
\textit{arXiv:1610.06328}

\bibitem{henkes05}
Henkes S, Chakraborty B. 2005.
Jamming as a critical phenomenon: A field theory of zero-temperature grain
  packings.
\textit{Phys. Rev. Lett.} 95:198002

\bibitem{edwards05}
Edwards S. 2005.
The full canonical ensemble of a granular system.
\textit{Physica A: Statistical Mechanics and its Applications} 353:114 -- 118

\bibitem{blumenfeld09}
Blumenfeld R, Edwards S. 2009.
On granular stress statistics: Compactivity, angoricity, and some open issues.
\textit{JOURNAL OF PHYSICAL CHEMISTRY B} 113:3981--3987

\bibitem{jenkins2016kinetic}
Jenkins JT. 2016.
Kinetic theories for collisional grain flows.
\textit{Handbook of Granular Materials} :155

\bibitem{PuckettD13}
Puckett JG, Daniels KE. 2013.
Equilibrating temperaturelike variables in jammed granular subsystems.
\textit{Phys. Rev. Lett.} 110:058001

\bibitem{LangerBL10}
Langer J, Bouchbinder E, Lookman T. 2010.
Thermodynamic theory of dislocation-mediated plasticity.
\textit{Acta Materialia} 58:3718--3732

\bibitem{MakseKurchan}
Makse HA, Kurchan J. 2002.
Testing the thermodynamic approach to granular matter with a numerical model of
  a decisive experiment.
\textit{Nature} 415:614--617

\bibitem{AbateD08}
Abate AR, Durian DJ. 2008.
Effective temperatures and activated dynamics for a two-dimensional air-driven
  granular system on two approaches to jamming.
\textit{Phys. Rev. Lett.} 101:245701

\bibitem{BerthierB02}
Berthier L, Barrat JL. 2002.
Shearing a glassy material: numerical tests of nonequilibrium mode-coupling
  approaches and experimental proposals.
\textit{Physical review letters} 89:095702

\bibitem{OnoODLLN02}
Ono IK, O'Hern CS, Durian DJ, Langer SA, Liu AJ, Nagel SR. 2002.
Effective temperatures of a driven system near jamming.
\textit{Phys. Rev. Lett.} 89:095703

\bibitem{jonason1998memory}
Jonason K, Vincent E, Hammann J, Bouchaud J, Nordblad P. 1998.
Memory and chaos effects in spin glasses.
\textit{Physical Review Letters} 81:3243

\bibitem{SethnaDKKRS93}
Sethna JP, Dahmen K, Kartha S, Krumhansl JA, Roberts BW, Shore JD. 1993.
Hysteresis and hierarchies - dynamics of disorder-driven 1st-order
  phase-transformations.
\textit{Physical Review Letters} 70:3347--3350

\bibitem{coppersmith1987simple}
Coppersmith S. 1987.
A simple illustration of "phase organization".
\textit{Physics Letters A} 125:473--475

\bibitem{coppersmith1987pulse}
Coppersmith S, Littlewood P. 1987.
Pulse-duration memory effect and deformable charge-density waves.
\textit{Physical Review B} 36:311

\bibitem{tang1987phase}
Tang C, Wiesenfeld K, Bak P, Coppersmith S, Littlewood P. 1987.
Phase organization.
\textit{Physical review letters} 58:1161

\bibitem{AsaroLubarda}
Asaro R, Lubarda V. 2006.
Mechanics of solids and materials.
Cambridge books online. Cambridge University Press

\bibitem{silva13}
de~Silva C. 2013.
Mechanics of materials.
Computational Mechanics and Applied Analysis. CRC Press

\bibitem{nair15}
Nair S. 2015.
Mechanics of aero-structures.
Cambridge University Press

\bibitem{philpot12}
Philpot T. 2012.
Mechanics of materials: An integrated learning system, 3rd edition: Third
  edition.
Wiley Global Education

\bibitem{MachtaCTS13}
Machta BB, Chachra R, Transtrum M, Sethna JP. 2013.
Parameter space compression underlies emergent theories and predictive models.
\textit{Science} 342:604--607

\bibitem{TranstrumMBDMS15}
Transtrum MK, Machta BB, Brown KS, Daniels BC, Myers CR, Sethna JP. 2015.
Perspective: Sloppiness and emergent theories in physics, biology, and beyond.
\textit{The Journal of Chemical Physics} 143:--

\bibitem{kocks2000texture}
Kocks UF, Tom{\'e} CN, Wenk HR. 2000.
Texture and anisotropy: preferred orientations in polycrystals and their effect
  on materials properties.
Cambridge university press

\bibitem{randle2000texture}
Randle V, Engler O. 2000.
Texture analysis.
\textit{Macrotexture, Microtexture and Orientation Mapping, Gordon and Breach,
  London}

\bibitem{bhattacharya03}
Bhattacharya K. 2003.
Microstructure of martensite: Why it forms and how it gives rise to the
  shape-memory effect.
Oxford University Press, Oxford

\bibitem{Corte08}
Cort{\'e} L, Chaikin PM, Gollub JP, Pine DJ. 2008.
Random organization in periodically driven systems.
\textit{Nature Physics} 4:420 -- 424

\bibitem{Pine05}
Pine D, Gollub J, Brady J, Leshansky A. 2005.
Chaos and threshold for irreversibility in sheared suspensions.
\textit{Nature} 438:997–1000

\bibitem{Paulsen14}
Paulsen JD, Keim NC, Nagel SR. 2014.
Multiple transient memories in experiments on sheared non-brownian suspensions.
\textit{Phys. Rev. Lett.} 113

\bibitem{Reichhardt09}
Reichhardt C, Reichhardt CJO. 2009.
Random organization and plastic depinning.
\textit{Phys. Rev. Lett.} 103

\bibitem{Keim13}
Keim NC, Arratia PE. 2013.
Yielding and microstructure in a 2d jammed material under shear deformation.
\textit{Soft Matter} 9:6222--6225

\bibitem{Keim14}
Keim NC, Arratia PE. 2014.
Mechanical and microscopic properties of the reversible plastic regime in a 2d
  jammed material.
\textit{Phys. Rev. Lett.} 112

\bibitem{Menon09}
Menon GI, Ramaswamy S. 2009.
Universality class of the reversible-irreversible transition in sheared
  suspensions.
\textit{Phys. Rev. E} 79

\bibitem{Regev15}
Regev I, Weber J, Reichhardt C, Dahmen KA, Lookman T. 2015.
Reversibility and criticality in amorphous solids.
\textit{Nature Communications} 6

\bibitem{Regev13}
Regev I, Lookman T, Reichhardt C. 2013.
Onset of irreversibility and chaos in amorphous solids under periodic shear.
\textit{Phys. Rev. E} 88

\bibitem{Fiocco13}
Fiocco D, Foffi G, Sastry S. 2013.
Oscillatory athermal quasistatic deformation of a model glass.
\textit{Phys. Rev. E} 88

\bibitem{Jeanneret14}
Jeanneret R, Bartolo D. 2014.
Geometrically protected reversibility in hydrodynamic loschmidt-echo
  experiments.
\textit{Nat. Commun.} 5

\bibitem{Nagamanasa14}
Nagamanasa KH, Gokhale S, Sood AK, Ganapathy R. 2014.
Experimental signatures of a nonequilibrium phase transition governing the
  yielding of a soft glass.
\textit{Phys. Rev. E Stat. Nonlin. Soft Matter Phys.} 89

\bibitem{Rogers14}
Rogers MC, Chen K, Andrzejewski L, Narayanan S, Ramakrishnan S, et~al. 2014.
Echoes in x-ray speckles track nanometer-scale plastic events in colloidal gels
  under shear.
\textit{Phys. Rev. E Stat. Nonlin. Soft Matter Phys.} 90

\bibitem{Mobius14}
M{\"o}bius R, Heussinger C. 2014.
(ir)reversibility in dense granular systems driven by oscillating forces.
\textit{Soft Matter} 10:4806–4812

\bibitem{Schreck13}
Schreck CF, Hoy RS, Shattuck MD, O'Hern CS. 2013.
Particle-scale reversibility in athermal particulate media below jamming.
\textit{Phys. Rev. E Stat. Nonlin. Soft Matter Phys.} 88

\bibitem{Slotterback12}
Slotterback Sea. 2012.
Onset of irreversibility in cyclic shear of granular packings.
\textit{Phys. Rev. E Stat. Nonlin. Soft Matter Phys.} 85

\bibitem{Royer15}
Royer JR, Chaikin PM. 2015.
Precisely cyclic sand: self-organization of periodically sheared frictional
  grains.
\textit{Proc. Natl Acad. Sci.} 112:49–53

\bibitem{Zhou14}
Zhou C, Olson~Reichhardt C, Reichhardt C, Beyerlein I. 2014.
Random organization in periodically driven gliding dislocations.
\textit{Phys. Lett. A} 378

\bibitem{Okuma11}
Okuma S, Tsugawa Y, Motohashi A. 2011.
Transition from reversible to irreversible flow: Absorbing and depinning
  transitions in a sheared-vortex system.
\textit{Phys. Rev. B} 83

\bibitem{Mangan08}
Mangan N, Reichhardt C, Reichhardt C. 2008.
Reversible to irreversible flow transition in periodically driven vortices.
\textit{Phys. Rev. Lett.} 100

\bibitem{Perez11}
P\'erez~Daroca D, Pasquini G, Lozano GS, Bekeris V. 2011.
Dynamics of superconducting vortices driven by oscillatory forces in the
  plastic-flow regime.
\textit{Phys. Rev. B} 84

\bibitem{Lopez99}
L{\'o}pez D, Kwok WK, Safar H, Olsson RJ, Petrean AM, et~al. 1999.
Spatially resolved dynamic correlation in the vortex state of high temperature
  superconductors.
\textit{Phys. Rev. Lett.} 82:1277--1280

\bibitem{Miguel03}
Miguel MC, Zapperi S. 2003.
Tearing transition and plastic flow in superconducting thin films.
\textit{Nature Mat.} 2

\bibitem{Shaw12}
Shaw G, Mandal P, Banerjee SS, Niazi A, Rastogi AK, et~al. 2012.
Critical behavior at depinning of driven disordered vortex matter in 2h-nbs2.
\textit{Phys. Rev. B} 85

\bibitem{Okuma12}
Okuma S, Motohashi A. 2012.
Critical behavior associated with transient dynamics near the depinning
  transition.
\textit{New J. Phys.} 14

\bibitem{goldenfeld1992lectures}
Goldenfeld N. 1992.
Lectures on phase transitions and the renormalization group

\bibitem{SethnaDM01}
Sethna JP, Dahmen KA, Myers CR. 2001.
Crackling noise.
\textit{Nature} 410:242--250

\bibitem{CsikorMWZZ04}
Csikor FF, Motz C, Weygand D, Zaiser M, Zapperi S. 2007{\natexlab{a}}.
Dislocation avalanches, strain bursts, and the problem of plastic forming at
  the micrometer scale.
\textit{Science} 318:251--254

\bibitem{friedman12}
Friedman N, Jennings AT, Tsekenis G, Kim JY, Tao M, et~al. 2012.
Statistics of dislocation slip avalanches in nanosized single crystals show
  tuned critical behavior predicted by a simple mean field model.
\textit{Phys. Rev. Lett.} 109:095507

\bibitem{BurridgeKnopoff}
Burridge R, Knopoff L. 1967.
Model and theoretical seismicity.
\textit{Bulletin of the seismological society of america} 57:341--371

\bibitem{bak1989earthquakes}
Bak P, Tang C. 1989.
Earthquakes as a self-organized critical phenomenon.
\textit{J. geophys. Res} 94:635--15

\bibitem{petri1994experimental}
Petri A, Paparo G, Vespignani A, Alippi A, Costantini M. 1994.
Experimental evidence for critical dynamics in microfracturing processes.
\textit{Physical review letters} 73:3423

\bibitem{garcimartin1997statistical}
Garcimartin A, Guarino A, Bellon L, Ciliberto S. 1997.
Statistical properties of fracture precursors.
\textit{Physical Review Letters} 79:3202

\bibitem{ChenPSZD11}
Chen Y, Papanikolaou S, Sethna JP, Zapperi S, Durin G. 2011.
Avalanche spatial structure and multivariable scaling functions; sizes,
  heights, widths, and views through windows.
\textit{Physical Review E} 84:061103

\bibitem{PapanikolaouBSDZS11}
Papanikolaou S, Bohn F, Sommer RL, Durin G, Zapperi S, Sethna JP. 2011.
Universality beyond power laws and the average avalanche shape.
\textit{Nature Physics} 7:316--320

\bibitem{Weiss}
Miguel MC, Vespignani A, Zapperi S, Weiss J, Grasso JR. 2001.
Intermittent dislocation flow in viscoplastic deformation.
\textit{Nature} 410:667--671

\bibitem{Dimiduk}
Dimiduk DM, Woodward C, LeSar R, Uchic MD. 2006.
Scale-free intermittent flow in crystal plasticity.
\textit{Science} 312:1188--1190

\bibitem{WeissM03}
Weiss J, Marsan D. 2003.
Three-dimensional mapping of dislocation avalanches: Clustering and space/time
  coupling.
\textit{Science} 299

\bibitem{mughrabi1983dislocation}
Mughrabi H. 1983.
Dislocation wall and cell structures and long-range internal stresses in
  deformed metal crystals.
\textit{Acta metallurgica} 31:1367--1379

\bibitem{hahner1998fractal}
H{\"a}hner P, Bay K, Zaiser M. 1998.
Fractal dislocation patterning during plastic deformation.
\textit{Physical review letters} 81:2470

\bibitem{hughes1998scaling}
Hughes D, Chrzan D, Liu Q, Hansen N. 1998.
Scaling of misorientation angle distributions.
\textit{Physical review letters} 81:4664

\bibitem{ChenCPS13}
Chen YS, Choi W, Papanikolaou S, Bierbaum M, Sethna JP. 2013.
Scaling theory of continuum dislocation dynamics in three dimensions:
  Self-organized fractal pattern formation.
\textit{International Journal of Plasticity} 46:94--129

\bibitem{ChenCPS10}
Chen YS, Choi W, Papanikolaou S, Sethna JP. 2010.
Bending crystals: the evolution of grain boundaries and fractal dislocation
  structures.
\textit{Phys. Rev. Lett.} 105:105501

\bibitem{BertalanSSZ14}
Bertalan Z, Shekhawat A, Sethna JP, Zapperi S. 2014.
Fracture strength: Stress concentration, extreme value statistics, and the fate
  of the weibull distribution.
\textit{Phys. Rev. Applied} 2:034008

\bibitem{JaronSSnn}
Kent-Dobias J, Shekhawat A, Sethna JP. 2016.
(work in progress)

\bibitem{alava2006morphology}
Alava MJ, Nukala PK, Zapperi S. 2006{\natexlab{a}}.
Morphology of two-dimensional fracture surfaces.
\textit{Journal of Statistical Mechanics: Theory and Experiment} 2006:L10002

\bibitem{zapperi2005crack}
Zapperi S, Nukala PKV, {\v{S}}imunovi{\'c} S. 2005.
Crack roughness and avalanche precursors in the random fuse model.
\textit{Physical Review E} 71:026106

\bibitem{alava2006statistical}
Alava MJ, Nukala PK, Zapperi S. 2006{\natexlab{b}}.
Statistical models of fracture.
\textit{Advances in Physics} 55:349--476

\bibitem{bouchaud1990fractal}
Bouchaud E, Lapasset G, Planes J. 1990.
Fractal dimension of fractured surfaces: a universal value?
\textit{EPL (Europhysics Letters)} 13:73

\bibitem{bouchaud1997scaling}
Bouchaud E. 1997.
Scaling properties of cracks.
\textit{Journal of Physics: Condensed Matter} 9:4319

\bibitem{ponson2006two}
Ponson L, Bonamy D, Bouchaud E. 2006.
Two-dimensional scaling properties of experimental fracture surfaces.
\textit{Physical review letters} 96:035506

\bibitem{morel2000scaling}
Morel S, Schmittbuhl J, Bouchaud E, Valentin G. 2000.
Scaling of crack surfaces and implications for fracture mechanics.
\textit{Physical review letters} 85:1678

\bibitem{maaloy1992experimental}
M{\aa}l{\o}y KJ, Hansen A, Hinrichsen EL, Roux S. 1992.
Experimental measurements of the roughness of brittle cracks.
\textit{Physical Review Letters} 68:213

\bibitem{hansen2003origin}
Hansen A, Schmittbuhl J. 2003.
Origin of the universal roughness exponent of brittle fracture surfaces:
  stress-weighted percolation in the damage zone.
\textit{Physical review letters} 90:045504

\bibitem{schmittbuhl2003roughness}
Schmittbuhl J, Hansen A, Batrouni GG. 2003.
Roughness of interfacial crack fronts: stress-weighted percolation in the
  damage zone.
\textit{Physical review letters} 90:045505

\bibitem{laurson2010avalanches}
Laurson L, Santucci S, Zapperi S. 2010.
Avalanches and clusters in planar crack front propagation.
\textit{Physical Review E} 81:046116

\bibitem{schmittbuhl1995interfacial}
Schmittbuhl J, Roux S, Vilotte JP, M{\aa}l{\o}y KJ. 1995.
Interfacial crack pinning: effect of nonlocal interactions.
\textit{Physical Review Letters} 74:1787

\bibitem{rosso2002roughness}
Rosso A, Krauth W. 2002.
Roughness at the depinning threshold for a long-range elastic string.
\textit{Physical Review E} 65:025101

\bibitem{mecholsky1989quantitative}
Mecholsky J, Passoja D, Feinberg-Ringel K. 1989.
Quantitative analysis of brittle fracture surfaces using fractal geometry.
\textit{Journal of the American Ceramic Society} 72:60--65

\bibitem{mecholsky1988self}
Mecholsky J, Mackin T, Passoja D. 1988.
Self-similar crack propagation in brittle materials.
\textit{Fractography of Glasses and Ceramics Westerville, Ohio, 1988,}
  22:127--134

\bibitem{mecholsky1991relationship}
Mecholsky JJ, Freiman SW. 1991.
Relationship between fractal geometry and fractography.
\textit{Journal of the American Ceramic Society} 74:3136--3138

\bibitem{tsai1991fractal}
Tsai Y, Mecholsky J. 1991.
Fractal fracture of single crystal silicon.
\textit{Journal of Materials Research} 6:1248--1263

\bibitem{schmittbuhl1997direct}
Schmittbuhl J, M{\aa}l{\o}y KJ. 1997.
Direct observation of a self-affine crack propagation.
\textit{Physical review letters} 78:3888

\bibitem{schmittbuhl1995scaling}
Schmittbuhl J, Schmitt F, Scholz C. 1995.
Scaling invariance of crack surfaces.
\textit{Journal of Geophysical Research: Solid Earth} 100:5953--5973

\bibitem{schmittbuhl1993field}
Schmittbuhl J, Gentier S, Roux S. 1993.
Field measurements of the roughness of fault surfaces.
\textit{Geophysical Research Letters} 20:639--641

\bibitem{santucci2007statistics}
Santucci S, M{\aa}l{\o}y KJ, Delaplace A, Mathiesen J, Hansen A, et~al. 2007.
Statistics of fracture surfaces.
\textit{Physical review E} 75:016104

\bibitem{ChenZS15}
Chen YJ, Zapperi S, Sethna JP. 2015.
Crossover behavior in interface depinning.
\textit{Phys. Rev. E} 92:022146

\bibitem{talreja2013probability}
Talreja R, Weibull W. 2013.
Probability of fatigue failure based on residual strength. In \textit{ICF4,
  Waterloo (Canada) 1977}

\bibitem{jayatilaka1977statistical}
Jayatilaka AdS, Trustrum K. 1977.
Statistical approach to brittle fracture.
\textit{Journal of Materials Science} 12:1426--1430

\bibitem{phoenix1973asymptotic}
Phoenix SL, Taylor HM. 1973.
The asymptotic strength distribution of a general fiber bundle.
\textit{Advances in Applied Probability} :200--216

\bibitem{gyorgyi2010renormalization}
Gy{\"o}rgyi G, Moloney N, Ozog{\'a}ny K, R{\'a}cz Z, Droz M. 2010.
Renormalization-group theory for finite-size scaling in extreme statistics.
\textit{Physical Review E} 81:041135

\bibitem{gyorgyi2008finite}
Gy{\"o}rgyi G, Moloney N, Ozog{\'a}ny K, R{\'a}cz Z. 2008.
Finite-size scaling in extreme statistics.
\textit{Physical review letters} 100:210601

\bibitem{salminen2002acoustic}
Salminen L, Tolvanen A, Alava MJ. 2002.
Acoustic emission from paper fracture.
\textit{Physical Review Letters} 89:185503

\bibitem{koivisto2007creep}
Koivisto J, Rosti J, Alava MJ. 2007.
Creep of a fracture line in paper peeling.
\textit{Physical review letters} 99:145504

\bibitem{hemmer1992distribution}
Hemmer PC, Hansen A. 1992.
The distribution of simultaneous fiber failures in fiber bundles.
\textit{Journal of applied mechanics} 59:909--914

\bibitem{ShekhawatZS13}
Shekhawat A, Zapperi S, Sethna JP. 2013.
From damage percolation to crack nucleation through finite-size criticality.
\textit{Physical Review Letters} 110:185505

\bibitem{fisher98}
Fisher DS. 1998.
Collective transport in random media: from superconductors to earthquakes.
\textit{Physics Reports} 301:113 -- 150

\bibitem{zaiser05}
Zaiser M, Moretti P. 2005.
Fluctuation phenomena in crystal plasticity—a continuum model.
\textit{Journal of Statistical Mechanics: Theory and Experiment} 2005:P08004

\bibitem{zaiser06}
Zaiser M. 2006.
Scale invariance in plastic flow of crystalline solids.
\textit{Advances in Physics} 55:185--245

\bibitem{talamali11}
Talamali M, Pet\"aj\"a V, Vandembroucq D, Roux S. 2011.
Avalanches, precursors, and finite-size fluctuations in a mesoscopic model of
  amorphous plasticity.
\textit{Phys. Rev. E} 84:016115

\bibitem{ZapperiDurinReview}
Durin G, Zapperi S. 2006.
Chapter 3 - the barkhausen effect. In \textit{The Science of Hysteresis}, eds.
  G~Bertotti, ID~Mayergoyz. Oxford: Academic Press,  181 -- 267

\bibitem{FunctionalRGMagnetism1}
Narayan O, Fisher DS. 1993.
Threshold critical dynamics of driven interfaces in random media.
\textit{Phys. Rev. B} 48:7030--7042

\bibitem{FunctionalRGMagnetism2}
Leschhorn H, Nattermann T, Stepanow S, Tang LH. 1997.
Driven interface depinning in a disordered medium.
\textit{Annalen der Physik} 509:1--34

\bibitem{FunctionalRGMagnetism3}
Chauve P, Giamarchi T, Le~Doussal P. 2000.
Creep and depinning in disordered media.
\textit{Phys. Rev. B} 62:6241--6267

\bibitem{FunctionalRGMagnetism4}
Chauve P, Le~Doussal P, J\"org~Wiese K. 2001.
Renormalization of pinned elastic systems: How does it work beyond one loop?
\textit{Phys. Rev. Lett.} 86:1785--1788

\bibitem{FunctionalRGMagnetism5}
Le~Doussal P, Wiese KJ, Chauve P. 2002.
Two-loop functional renormalization group theory of the depinning transition.
\textit{Phys. Rev. B} 66:174201

\bibitem{DahmenEarthquakes3}
Kagan YY. 2010.
Earthquake size distribution: Power-law with exponent $\beta=1/2$?
\textit{Tectonophysics} 490:103 -- 114

\bibitem{DahmenEarthquakes1}
Ben-Zion Y. 2008.
Collective behavior of earthquakes and faults: Continuum-discrete transitions,
  progressive evolutionary changes, and different dynamic regimes.
\textit{Reviews of Geophysics} 46:n/a--n/a.
RG4006

\bibitem{BaroCIPSSSV13}
Bar\'o J, Corral A, Illa X, Planes A, Salje EKH, et~al. 2013.
Statistical similarity between the compression of a porous material and
  earthquakes.
\textit{Phys. Rev. Lett.} 110:088702

\bibitem{FisherDRB97}
Fisher DS, Dahmen KA, Ramanathan D, Ben-Zion Y. 1997.
Statistics of earthquakes in simple models of heterogeneous faults.
\textit{Phys. Rev. Lett.} 78:4885--4888

\bibitem{chen91}
Chen K, Bak P, Obukhov SP. 1991.
Self-organized criticality in a crack-propagation model of earthquakes.
\textit{Phys. Rev. A} 43:625--630

\bibitem{DahmenEarthquakes4}
Carlson JM, Langer JS. 1989.
Properties of earthquakes generated by fault dynamics.
\textit{Phys. Rev. Lett.} 62:2632--2635

\bibitem{MyersCarlsonEarthquakes}
Langer J, Carlson J, Myers CR, Shaw BE. 1996.
Slip complexity in dynamic models of earthquake faults.
\textit{Proceedings of the National Academy of Sciences} 93:3825--3829

\bibitem{ErtasD}
Dahmen K, Erta\ifmmode~\mbox{\c{s}}\else \c{s}\fi{} D, Ben-Zion Y. 1998.
Gutenberg-richter and characteristic earthquake behavior in simple mean-field
  models of heterogeneous faults.
\textit{Phys. Rev. E} 58:1494--1501

\bibitem{liu10}
Liu AJ, Nagel SR. 2010.
The jamming transition and the marginally jammed solid.
\textit{Annual Review of Condensed Matter Physics} 1:347--369

\bibitem{liu11}
Liu AJ, Nagel SR, van Saarloos W, Wyart M. 2011.
The jamming scenario---an introduction and outlook.
Oxford Scholarship Online: September 2011,  1--72

\bibitem{ispanovity14}
Isp\'anovity PD, Laurson L, Zaiser M, Groma I, Zapperi S, Alava MJ. 2014.
Avalanches in 2d dislocation systems: Plastic yielding is not depinning.
\textit{Phys. Rev. Lett.} 112:235501

\bibitem{bi16}
Bi D, Yang X, Marchetti MC, Manning ML. 2016.
Motility-driven glass and jamming transitions in biological tissues.
\textit{Phys. Rev. X} 6:021011

\bibitem{tsekenis11}
Tsekenis G, Goldenfeld N, Dahmen KA. 2011.
Dislocations jam at any density.
\textit{Phys. Rev. Lett.} 106:105501

\bibitem{miguel02}
Miguel MC, Vespignani A, Zaiser M, Zapperi S. 2002.
Dislocation jamming and andrade creep.
\textit{Phys. Rev. Lett.} 89:165501

\bibitem{GoodrichLS16}
Goodrich CP, Liu AJ, Sethna JP. 2016.
Scaling ansatz for the jamming transition.
\textit{Proceedings of the National Academy of Sciences} 113

\bibitem{hatano08}
Hatano T. 2008.
Scaling properties of granular rheology near the jamming transition.
\textit{Journal of the Physical Society of Japan} 77:123002

\bibitem{tighe10}
Tighe BP, Woldhuis E, Remmers JJC, van Saarloos W, van Hecke M. 2010.
Model for the scaling of stresses and fluctuations in flows near jamming.
\textit{Phys. Rev. Lett.} 105:088303

\bibitem{dinkgreve15}
Dinkgreve M, Paredes J, Michels MAJ, Bonn D. 2015.
Universal rescaling of flow curves for yield-stress fluids close to jamming.
\textit{Phys. Rev. E} 92:012305

\bibitem{SuperPlastic}
Nieh TG, Wadsworth J, Sherby OD. 2005.
Superplasticity in metals and ceramics.
Cambridge university press

\bibitem{Pazmandi:2000ep}
P{\'a}zm{\'a}ndi F, Zar{\'a}nd G, Zim{\'a}nyi GT. 2000.
{Self-organized criticality in the hysteresis of the
  Sherrington{\textendash}Kirkpatrick model}.
\textit{Physica B: Condensed Matter} 275:207--211

\bibitem{WyartMeanField}
Lin J, Wyart M. 2016.
Mean-field description of plastic flow in amorphous solids.
\textit{Phys. Rev. X} 6:011005

\bibitem{Dahmen:2009iz}
Dahmen KA, Ben-Zion Y, Uhl JT. 2009.
{Micromechanical Model for Deformation in Solids with Universal Predictions for
  Stress-Strain Curves and Slip Avalanches}.
\textit{Physical Review Letters} 102:175501

\bibitem{Sun:2010bh}
Sun BA, Yu HB, Jiao W, Bai HY, Zhao DQ, Wang WH. 2010.
{Plasticity of Ductile Metallic Glasses: A Self-Organized Critical State}.
\textit{Physical Review Letters} 105:035501

\bibitem{Antonaglia:2014gc}
Antonaglia J, Wright WJ, Gu X, Byer RR, Hufnagel TC, et~al. 2014.
{Bulk Metallic Glasses Deform via Slip Avalanches}.
\textit{Physical Review Letters} 112:155501

\bibitem{Liu:2016ft}
Liu C, Ferrero EE, Puosi F, Barrat JL, Martens K. 2016.
{Driving Rate Dependence of Avalanche Statistics and Shapes at the Yielding
  Transition}.
\textit{Physical Review Letters} 116:065501

\bibitem{Budrikis:2015tu}
Budrikis Z, Fernandez-Castellanos D, Sandfeld S, Zaiser M, Zapperi S. 2015.
{Universality of Avalanche Exponents in Plastic Deformation of Disordered
  Solids}

\bibitem{Salerno:2013dz}
Salerno KM, Robbins MO. 2013.
{Effect of inertia on sheared disordered solids: Critical scaling of avalanches
  in two and three dimensions}.
\textit{Phys. Rev. E} 88:062206--15

\bibitem{SOCReview}
Bak P, Tang C, Wiesenfeld K. 1988.
Self-organized criticality.
\textit{Phys. Rev. A} 38:364--374

\bibitem{rollett2004recrystallization}
Rollett A, Humphreys F, Rohrer GS, Hatherly M. 2004.
Recrystallization and related annealing phenomena.
Elsevier

\bibitem{csikor2007dislocation}
Csikor FF, Motz C, Weygand D, Zaiser M, Zapperi S. 2007{\natexlab{b}}.
Dislocation avalanches, strain bursts, and the problem of plastic forming at
  the micrometer scale.
\textit{Science} 318:251--254

\bibitem{zaiser2005yielding}
Zaiser M, Marmo B, Moretti P. 2005.
The yielding transition in crystal plasticity--discrete dislocations and
  continuum models. In \textit{International Conference on Statistical
  Mechanics of Plasticity and Related Instabilities, Indian Institute of
  Science, Bangalore, India}

\bibitem{Simulations}
Lehtinen A, Costantini G, Alava MJ, Zapperi S, Laurson L. 2016.
Glassy features of crystal plasticity.
\textit{Phys. Rev. B} 94:064101

\bibitem{PapanikolaouDCSUWZ12}
Papanikolaou S, Dimiduk DM, Choi W, Sethna JP, Uchic MD, et~al. 2012.
Quasi-periodic events in crystal plasticity and the self-organized avalanche
  oscillator.
\textit{Nature} 490:517--521

\bibitem{jagla14}
Jagla EA, Landes FmcP, Rosso A. 2014.
Viscoelastic effects in avalanche dynamics: A key to earthquake statistics.
\textit{Phys. Rev. Lett.} 112:174301

\bibitem{ZimanyiSpinGlass}
P\'azm\'andi F, Zar\'and G, Zim\'anyi GT. 1999.
Self-organized criticality in the hysteresis of the sherrington-kirkpatrick
  model.
\textit{Phys. Rev. Lett.} 83:1034--1037

\bibitem{RutenbergAnisotropic}
Rutenberg AD, Vollmayr-Lee BP. 1999.
Anisotropic coarsening: Grain shapes and nonuniversal persistence.
\textit{Phys. Rev. Lett.} 83:3772--3775

\bibitem{ShoreHS92}
Shore JD, Holzer M, Sethna JP. 1992.
Logarithmically slow domain growth in nonrandomly frustrated systems -
  {I}sing-models with competing interactions.
\textit{Physical Review B} 46:11376--11404

\end{thebibliography}
\end{document}